\pdfoutput=1

\documentclass[12pt,a4paper]{article}

\usepackage{ifthen} 
\newboolean{pdflatex}
\setboolean{pdflatex}{true} 

\newboolean{articletitles}
\setboolean{articletitles}{true} 

\newboolean{uprightparticles}
\setboolean{uprightparticles}{false} 

\newboolean{inbibliography}
\setboolean{inbibliography}{false} 

\usepackage{microtype}
\usepackage{lineno}  
\usepackage{xspace} 
\usepackage{caption} 

\usepackage{graphicx}  
\usepackage{color}
\usepackage{colortbl}
\graphicspath{{./figs/}} 

\usepackage{amsmath} 
\usepackage{amssymb}
\usepackage{amsfonts}
\usepackage{upgreek} 

\newcommand*\patchAmsMathEnvironmentForLineno[1]{%
\expandafter\let\csname old#1\expandafter\endcsname\csname #1\endcsname
\expandafter\let\csname oldend#1\expandafter\endcsname\csname
end#1\endcsname
 \renewenvironment{#1}%
   {\linenomath\csname old#1\endcsname}%
   {\csname oldend#1\endcsname\endlinenomath}%
}
\newcommand*\patchBothAmsMathEnvironmentsForLineno[1]{%
  \patchAmsMathEnvironmentForLineno{#1}%
  \patchAmsMathEnvironmentForLineno{#1*}%
}
\AtBeginDocument{%
\patchBothAmsMathEnvironmentsForLineno{equation}%
\patchBothAmsMathEnvironmentsForLineno{align}%
\patchBothAmsMathEnvironmentsForLineno{flalign}%
\patchBothAmsMathEnvironmentsForLineno{alignat}%
\patchBothAmsMathEnvironmentsForLineno{gather}%
\patchBothAmsMathEnvironmentsForLineno{multline}%
\patchBothAmsMathEnvironmentsForLineno{eqnarray}%
}

\usepackage{hyperref}    
\usepackage[all]{hypcap} 

\usepackage{xspace} 
\usepackage{upgreek}

\def\lhcb {\mbox{LHCb}\xspace}

\def\babar  {\mbox{BaBar}\xspace}
\def\belle  {\mbox{Belle}\xspace}

\def\MagUp {\mbox{\em Mag\kern -0.05em Up}\xspace}

\ifthenelse{\boolean{uprightparticles}}%
{

 \def\Ppi         {\ensuremath{\uppi}\xspace}                 
                  
 \def\Prho        {\ensuremath{\uprho}\xspace}

 \def\PDelta      {\ensuremath{\Delta}\xspace}                 
 \def\PXi      {\ensuremath{\Xi}\xspace}                 
 \def\PLambda      {\ensuremath{\Lambda}\xspace}                 
 \def\PSigma      {\ensuremath{\Sigma}\xspace}                 
 \def\POmega      {\ensuremath{\Omega}\xspace}                 
 \def\PUpsilon      {\ensuremath{\Upsilon}\xspace}                 
 
 \def\PB      {\ensuremath{\mathrm{B}}\xspace}                 
                  
 \def\PD      {\ensuremath{\mathrm{D}}\xspace}

 \def\PK      {\ensuremath{\mathrm{K}}\xspace}

 \def\Pb      {\ensuremath{\mathrm{b}}\xspace}                 
 \def\Pc      {\ensuremath{\mathrm{c}}\xspace}

 \def\Ph      {\ensuremath{\mathrm{h}}\xspace}                 
 \def\Pi      {\ensuremath{\mathrm{i}}\xspace}

 \def\Ps      {\ensuremath{\mathrm{s}}\xspace}

}
{

 \def\Ppi         {\ensuremath{\pi}\xspace}                 
                  
 \def\Prho        {\ensuremath{\rho}\xspace}

 \mathchardef\PDelta="7101
 \mathchardef\PXi="7104
 \mathchardef\PLambda="7103
 \mathchardef\PSigma="7106
 \mathchardef\POmega="710A
 \mathchardef\PUpsilon="7107
                  
 \def\PB      {\ensuremath{B}\xspace}                 
                  
 \def\PD      {\ensuremath{D}\xspace}

 \def\PK      {\ensuremath{K}\xspace}

 \def\Pb      {\ensuremath{b}\xspace}                 
 \def\Pc      {\ensuremath{c}\xspace}

 \def\Ph      {\ensuremath{h}\xspace}                 
 \def\Pi      {\ensuremath{i}\xspace}

 \def\Ps      {\ensuremath{s}\xspace}

}

\makeatletter
\ifcase \@ptsize \relax
  \newcommand{\miniscule}{\@setfontsize\miniscule{4}{5}}
\or
  \newcommand{\miniscule}{\@setfontsize\miniscule{5}{6}}
\or
  \newcommand{\miniscule}{\@setfontsize\miniscule{5}{6}}
\fi
\makeatother

\DeclareRobustCommand{\optbar}[1]{\shortstack{{\miniscule (\rule[.5ex]{1.25em}{.18mm})}
  \\ [-.7ex] $#1$}}

\def\squark    {{\ensuremath{\Ps}}\xspace}

\def\cquark    {{\ensuremath{\Pc}}\xspace}

\def\bquark    {{\ensuremath{\Pb}}\xspace}

\def\pion   {{\ensuremath{\Ppi}}\xspace}
\def\piz    {{\ensuremath{\pion^0}}\xspace}

\def\pip    {{\ensuremath{\pion^+}}\xspace}
\def\pim    {{\ensuremath{\pion^-}}\xspace}
\def\pipm   {{\ensuremath{\pion^\pm}}\xspace}

\def\rhomeson {{\ensuremath{\Prho}}\xspace}
\def\rhoz     {{\ensuremath{\rhomeson^0}}\xspace}

\def\kaon    {{\ensuremath{\PK}}\xspace}
  \def\Kbar    {{\kern 0.2em\overline{\kern -0.2em \PK}{}}\xspace}

\def\KorKbar    {\kern 0.18em\optbar{\kern -0.18em K}{}\xspace}
\def\Kz      {{\ensuremath{\kaon^0}}\xspace}
\def\Kzb     {{\ensuremath{\Kbar{}^0}}\xspace}
\def\Kp      {{\ensuremath{\kaon^+}}\xspace}
\def\Km      {{\ensuremath{\kaon^-}}\xspace}
\def\Kpm     {{\ensuremath{\kaon^\pm}}\xspace}

\def\KS      {{\ensuremath{\kaon^0_{\mathrm{ \scriptscriptstyle S}}}}\xspace}

\def\Kstarz  {{\ensuremath{\kaon^{*0}}}\xspace}
\def\Kstarzb {{\ensuremath{\Kbar{}^{*0}}}\xspace}
\def\Kstar   {{\ensuremath{\kaon^*}}\xspace}

  \def\Dbar    {{\kern 0.2em\overline{\kern -0.2em \PD}{}}\xspace}
\def\D       {{\ensuremath{\PD}}\xspace}

\def\DorDbar    {\kern 0.18em\optbar{\kern -0.18em D}{}\xspace}
\def\Dz      {{\ensuremath{\D^0}}\xspace}
\def\Dzb     {{\ensuremath{\Dbar{}^0}}\xspace}

\def\Dstar   {{\ensuremath{\D^*}}\xspace}

\def\Dstarz  {{\ensuremath{\D^{*0}}}\xspace}

\def\Dstarm  {{\ensuremath{\D^{*-}}}\xspace}
\def\Dstarpm {{\ensuremath{\D^{*\pm}}}\xspace}

\def\B       {{\ensuremath{\PB}}\xspace}
\def\Bbar    {{\ensuremath{\kern 0.18em\overline{\kern -0.18em \PB}{}}}\xspace}

\def\BorBbar    {\kern 0.18em\optbar{\kern -0.18em B}{}\xspace}
\def\Bz      {{\ensuremath{\B^0}}\xspace}
\def\Bzb     {{\ensuremath{\Bbar{}^0}}\xspace}

\def\Bub     {{\ensuremath{\B^-}}\xspace}

\def\Bm      {{\ensuremath{\Bub}}\xspace}
\def\Bpm     {{\ensuremath{\B^\pm}}\xspace}

\def\Bs      {{\ensuremath{\B^0_\squark}}\xspace}
\def\Bsb     {{\ensuremath{\Bbar{}^0_\squark}}\xspace}

  \def\Y#1S{\ensuremath{\PUpsilon{(#1S)}}\xspace}

\def\Lbar        {{\ensuremath{\kern 0.1em\overline{\kern -0.1em\PLambda}}}\xspace}
\def\LorLbar    {\kern 0.18em\optbar{\kern -0.18em \PLambda}{}\xspace}

\def\to                 {\ensuremath{\rightarrow}\xspace}

\def\CP                {{\ensuremath{C\!P}}\xspace}

\def\AT#1     {\ensuremath{A_{\mathrm{T}}^{#1}}\xspace}           

\def\C#1      {\ensuremath{\mathcal{C}_{#1}}\xspace}                       
\def\Cp#1     {\ensuremath{\mathcal{C}_{#1}^{'}}\xspace}                    
\def\Ceff#1   {\ensuremath{\mathcal{C}_{#1}^{\mathrm{(eff)}}}\xspace}        
\def\Cpeff#1  {\ensuremath{\mathcal{C}_{#1}^{'\mathrm{(eff)}}}\xspace}       
\def\Ope#1    {\ensuremath{\mathcal{O}_{#1}}\xspace}                       
\def\Opep#1   {\ensuremath{\mathcal{O}_{#1}^{'}}\xspace}                    

\newcommand{\tev}{\ifthenelse{\boolean{inbibliography}}{\ensuremath{~T\kern -0.05em eV}\xspace}{\ensuremath{\mathrm{\,Te\kern -0.1em V}}}\xspace}
\newcommand{\gev}{\ensuremath{\mathrm{\,Ge\kern -0.1em V}}\xspace}
\newcommand{\mev}{\ensuremath{\mathrm{\,Me\kern -0.1em V}}\xspace}
\newcommand{\kev}{\ensuremath{\mathrm{\,ke\kern -0.1em V}}\xspace}
\newcommand{\ev}{\ensuremath{\mathrm{\,e\kern -0.1em V}}\xspace}
\newcommand{\gevc}{\ensuremath{{\mathrm{\,Ge\kern -0.1em V\!/}c}}\xspace}
\newcommand{\mevc}{\ensuremath{{\mathrm{\,Me\kern -0.1em V\!/}c}}\xspace}
\newcommand{\gevcc}{\ensuremath{{\mathrm{\,Ge\kern -0.1em V\!/}c^2}}\xspace}
\newcommand{\gevgevcccc}{\ensuremath{{\mathrm{\,Ge\kern -0.1em V^2\!/}c^4}}\xspace}
\newcommand{\mevcc}{\ensuremath{{\mathrm{\,Me\kern -0.1em V\!/}c^2}}\xspace}

\def\mum  {\ensuremath{{\,\upmu\mathrm{m}}}\xspace}

\def\invfb   {\ensuremath{\mbox{\,fb}^{-1}}\xspace}

\newcommand{\chisq}{\ensuremath{\chi^2}\xspace}

\newcommand{\chisqip}{\ensuremath{\chi^2_{\text{IP}}}\xspace}

\def\gsim{{~\raise.15em\hbox{$>$}\kern-.85em
          \lower.35em\hbox{$\sim$}~}\xspace}
\def\lsim{{~\raise.15em\hbox{$<$}\kern-.85em
          \lower.35em\hbox{$\sim$}~}\xspace}

\def\sqs   {\ensuremath{\protect\sqrt{s}}\xspace}

\def\ptot       {\mbox{$p$}\xspace}
\def\pt         {\mbox{$p_{\mathrm{ T}}$}\xspace}

\def\degrees{\ensuremath{^{\circ}}\xspace}

\def\evtgen     {\mbox{\textsc{EvtGen}}\xspace}

\def\geant      {\mbox{\textsc{Geant4}}\xspace}

\def\photos     {\mbox{\textsc{Photos}}\xspace}

\def\pythia     {\mbox{\textsc{Pythia}}\xspace}

\def\tell1  {TELL1\xspace}
\def\ukl1   {UKL1\xspace}

\newcommand{\ie}{\mbox{\itshape i.e.}\xspace}

\newcommand{\BtoDKst}{\ensuremath{\Bz\to\PD\Kstarz}\xspace}
\newcommand{\BstoDKst}{\ensuremath{\Bs\to\PD\Kstarzb}\xspace}
\newcommand{\BtoDK}{\ensuremath{\PB\to\PD\PK}\xspace}
\newcommand{\BtoDstKst}{\ensuremath{\Bz\to\Dstarz\Kstarz}\xspace}
\newcommand{\BstoDstKst}{\ensuremath{\Bs\to\Dstarz\Kstarzb}\xspace}
\newcommand{\BtoDrho}{\ensuremath{\Bz\to\PD\rhoz}\xspace}

\newcommand{\BpmtoDKpm}{\ensuremath{\Bpm\to\PD\Kpm}\xspace}
\newcommand{\BmtoDKm}{\ensuremath{\Bm\to\PD\Km}\xspace}
\newcommand{\BztoDstmu}{\ensuremath{\Bz\to\Dstarm\mu^+\nu_\mu}\xspace}

\newcommand{\BztoDstmuX}{\ensuremath{\Bz\to\Dstarm\mu^+\nu_\mu X}\xspace}

\newcommand{\KsPiPi}{\ensuremath{\KS\pip\pim}\xspace}
\newcommand{\KsKK}{\ensuremath{\KS\Kp\Km}\xspace}
\newcommand{\Kshh}{\ensuremath{\KS\Ph^+\Ph^-}\xspace}

\newcommand{\DtoKsPiPi}{\ensuremath{\PD\to\KsPiPi}\xspace}
\newcommand{\DtoKsKK}{\ensuremath{\PD\to\KsKK}\xspace}
\newcommand{\DtoKshh}{\ensuremath{\PD\to\Kshh}\xspace}

\newcommand{\etaDst}{\ensuremath{\eta_{\Dstar\mu}}\xspace}
\newcommand{\etaDKst}{\ensuremath{\eta_{DK^{*0}}}\xspace}
\newcommand{\xpm}{\ensuremath{x_{\pm}}\xspace}
\newcommand{\ypm}{\ensuremath{y_{\pm}}\xspace}
\newcommand{\xp}{\ensuremath{x_{+}}\xspace}
\newcommand{\yp}{\ensuremath{y_{+}}\xspace}
\newcommand{\xm}{\ensuremath{x_{-}}\xspace}
\newcommand{\ym}{\ensuremath{y_{-}}\xspace}

\newcommand{\xy}{\ensuremath{\xpm, \ypm}\xspace}

\newcommand{\msqmin}{\ensuremath{m^2_-}\xspace}
\newcommand{\msqplus}{\ensuremath{m^2_+}\xspace}

\providecommand{\e}[1]{\ensuremath{\times 10^{#1}}}

\usepackage{cite} 
\usepackage{mciteplus}

\begin{document}

\renewcommand{\thefootnote}{\fnsymbol{footnote}}
\setcounter{footnote}{1}


\begin{titlepage}
\pagenumbering{roman}

\vspace*{-1.5cm}
\centerline{\large EUROPEAN ORGANIZATION FOR NUCLEAR RESEARCH (CERN)}
\vspace*{1.5cm}
\noindent
\begin{tabular*}{\linewidth}{lc@{\extracolsep{\fill}}r@{\extracolsep{0pt}}}
\ifthenelse{\boolean{pdflatex}}
{\vspace*{-2.7cm}\mbox{\!\!\!\includegraphics[width=.14\textwidth]{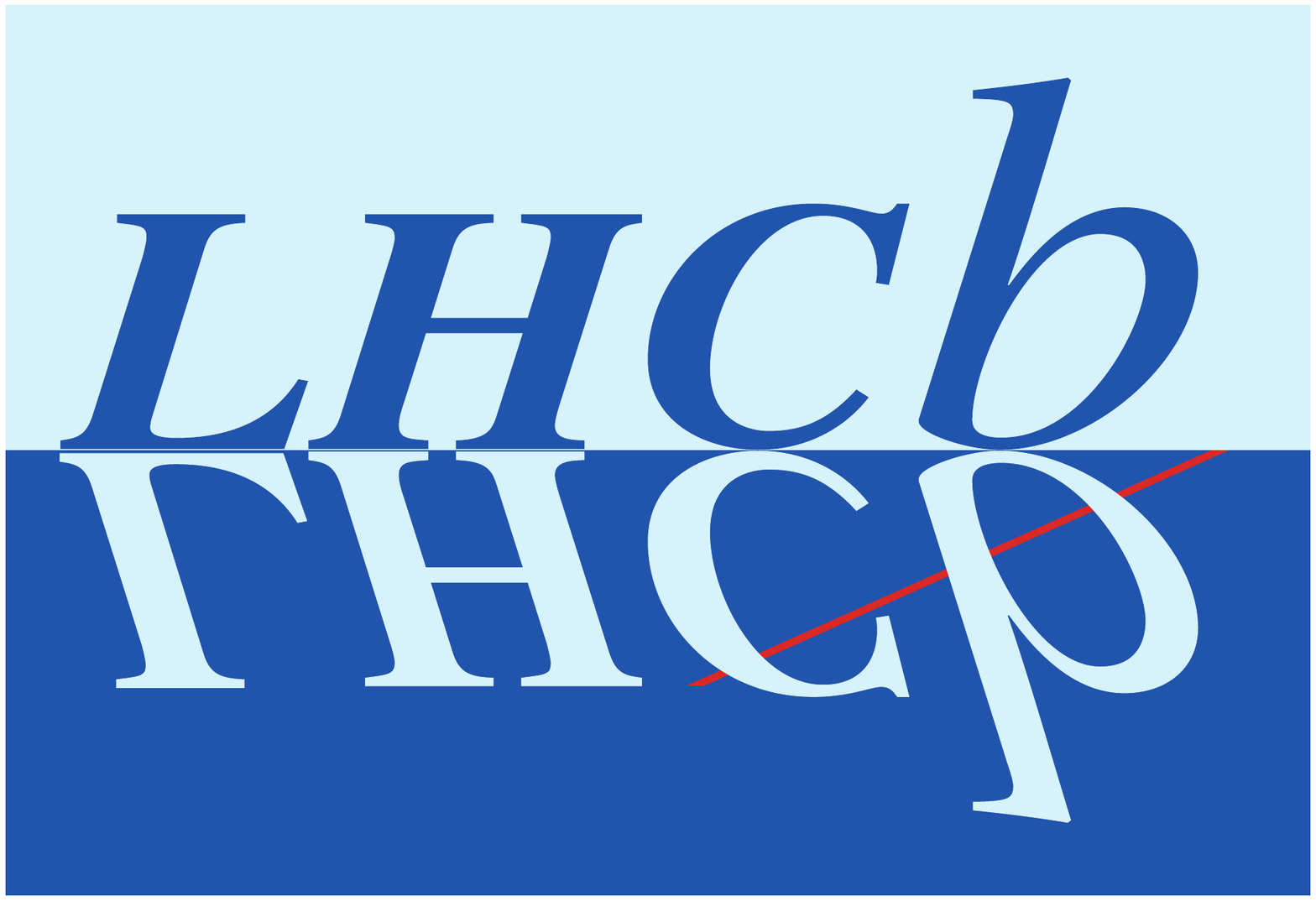}} & &}%
{\vspace*{-1.2cm}\mbox{\!\!\!\includegraphics[width=.12\textwidth]{lhcb-logo.eps}} & &}%
\\
 & & CERN-EP-2016-083 \\  
 & & LHCb-PAPER-2016-006 \\  
 & & 28 June 2016 \\ 
\end{tabular*}

\vspace*{1.0cm}

{\normalfont\bfseries\boldmath\huge
\begin{center}
  Model-independent measurement of the CKM angle $\gamma$ using $B^0 \to D\Kstarz$ decays with $D \to \KsPiPi$ and $\KsKK$ 
\end{center}
}

\begin{center}
The LHCb collaboration\footnote{Authors are listed at the end of this paper.}
\end{center}

\vspace{\fill}

\begin{abstract}
  \noindent
A binned Dalitz plot analysis of the decays \BtoDKst, with \DtoKsPiPi and \DtoKsKK, is performed to measure the observables $x_\pm$ and $y_\pm$, which are related to the CKM angle $\gamma$ and the hadronic parameters of the decays. The $D$ decay strong phase variation over the Dalitz plot is taken from measurements performed at the CLEO-c experiment, making the analysis independent of the $D$ decay model. With a sample of proton-proton collision data, corresponding to an integrated luminosity of 3.0 \invfb, collected by the LHCb experiment, the values of the \CP violation parameters are found to be $x_+ = 0.05 \pm 0.35 \pm 0.02$, $x_-=-0.31\pm 0.20 \pm 0.04$, $y_+=-0.81\pm 0.28\pm 0.06$ and $y_-=0.31\pm 0.21 \pm 0.05$, where the first uncertainties are statistical and the second systematic.
These observables correspond to values $\gamma$ = $(71 \pm 20)^\circ$, $r_{B^0} = 0.56\pm 0.17$ and $\delta_{B^0} = (204\,^{+21}_{-20})^\circ$. 
The parameters $r_{B^0}$ and $\delta_{B^0}$ are the magnitude ratio and strong phase difference between the suppressed and favoured $B^0$ decay amplitudes, and have been measured in a region of $\pm 50\mevcc$ around the $\Kstar(892)^{0}$ mass and with the magnitude of the cosine of the $\Kstar(892)^{0}$ helicity angle larger than 0.4. 

\end{abstract}

\begin{center}
  Published in JHEP 06 (2016) 131
\end{center}

\vspace{\fill}

{\footnotesize 
\centerline{\copyright~CERN on behalf of the \lhcb collaboration, licence \href{http://creativecommons.org/licenses/by/4.0/}{CC-BY-4.0}.}}
\vspace*{2mm}

\end{titlepage}


\newpage
\setcounter{page}{2}
\mbox{~}
%
%
%
%

\cleardoublepage


\renewcommand{\thefootnote}{\arabic{footnote}}
\setcounter{footnote}{0}


\cleardoublepage


\pagestyle{plain} 
\setcounter{page}{1}
\pagenumbering{arabic}


\section{Introduction}
\label{sec:Introduction}

The Standard Model (SM) description of \CP violation can be tested through measurements of the angle $\gamma$ of the unitarity triangle of the Cabibbo-Kobayashi-Maskawa (CKM) matrix~\cite{Cabibbo:1963yz,Kobayashi:1973fv}, where $\gamma \equiv \mathrm{arg} (-V_{ud}^{\phantom{\ast}}V_{ub}^\ast/V_{cd}^{\phantom{\ast}}V_{cb}^\ast)$. It is the only CKM angle easily accessible in tree-level processes and can be measured, with a small uncertainty from theory of $\delta\gamma/\gamma \leq 10^{-7}$~\cite{BrodZup}. Hence, in the absence of new physics effects at tree level~\cite{LENZ}, a precision measurement of $\gamma$ provides an SM benchmark which can be compared with other CKM matrix observables that are more likely to be affected by physics beyond the SM. Such comparisons are currently limited by the uncertainty on direct measurements of $\gamma$, which is about $7\degrees$~\cite{CKMFitter,UTFit}.

The CKM angle $\gamma$ is experimentally accessible through the interference between $\bar{b}\to\bar{c}u\bar{s}$ and $\bar{b}\to\bar{u}c\bar{s}$ transitions. The traditional golden mode is $\Bm\to\D\Km$, with charge-conjugation implied throughout, where $D$ represents a neutral $D$ meson reconstructed in a final state that is common to both \Dz and \Dzb decays. This mode has been studied at LHCb with a wide range of $D$ meson final states to measure observables with sensitivity to $\gamma$~\cite{LHCb-PAPER-2013-068,LHCb-PAPER-2014-041,LHCb-PAPER-2015-014,LHCb-PAPER-2016-003}. In addition to these studies, other $B$ decays have also been used with a variety of techniques to determine $\gamma$~\cite{LHCb-PAPER-2014-028, LHCb-PAPER-2014-038, LHCb-PAPER-2015-020, LHCb-PAPER-2015-059}.

\begin{figure}[bp]
\begin{center}
\includegraphics[width=0.8\textwidth]{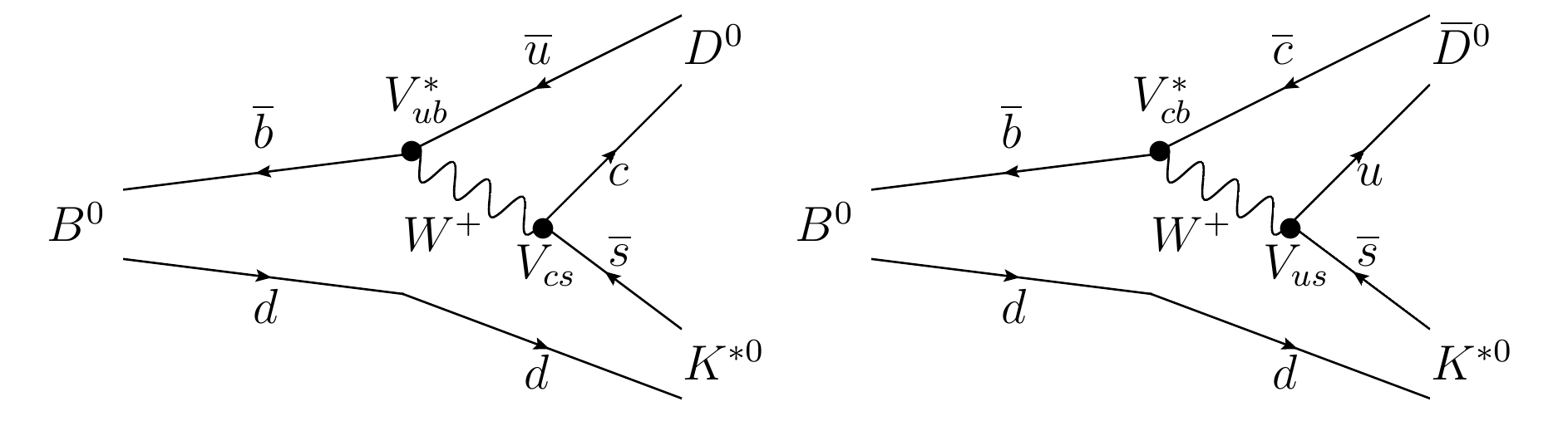}
\end{center}
\caption{Feynman diagrams of the (left) $\Bz \to \Dz\Kstarz$ and (right) $\Bz \to \Dzb\Kstarz$ amplitudes, which interfere in the \BtoDKst decay.}
\label{fig:feynman_diag}
\end{figure}

This paper presents an analysis in which the decay \BtoDKst provides sensitivity to the CKM angle $\gamma$ through the interfering amplitudes shown in Fig.~\ref{fig:feynman_diag}. Here the \Kstarz refers to the $\Kstar(892)^0$, and the charge of the kaon from the \Kstarz unambiguously identifies the flavour of the decaying \B meson as \Bz or \Bzb. Although the branching fraction of the \BtoDKst decay is an order of magnitude smaller than that of the $\Bm\to D\Km$ decay~\cite{PDG2014}, it is expected to exhibit larger \CP-violating effects as the two colour-suppressed Feynman diagrams in Fig.~\ref{fig:feynman_diag} are comparable in magnitude. Measurements sensitive to $\gamma$ using the \BtoDKst decay mode were pioneered by the \babar~\cite{bab2body} and \belle~\cite{belle2body} collaborations, and have been pursued by the LHCb collaboration~\cite{LHCb-PAPER-2014-028, LHCb-PAPER-2015-059}.

The three-body self-conjugate decays \DtoKsPiPi and \DtoKsKK, designated collectively as \DtoKshh, are accessible to both \Dz and \Dzb. They have large variation of the strong phase over the Dalitz plot, and thus provide a powerful method to determine the angle $\gamma$. Sensitivity to $\gamma$ is obtained by comparing the distribution of events in the \DtoKshh Dalitz plots of $B$ mesons reconstructed in each flavour, as described in Refs.~\cite{ads,GGSZ,BONDARGGSZ}. 
To determine $\gamma$ from the comparison, input is required on the variation within the Dalitz plot of the strong-interaction phase difference between \Dz and \Dzb decays. 
An amplitude model of the $\Dz\to\Kshh$ decay can be used to provide this information and this technique has been used to study the \BtoDKst, \DtoKsPiPi decay mode by \babar~\cite{BBGGSZ} and LHCb~\cite{LHCb-PAPER-2016-007}. In Ref.~\cite{LHCb-PAPER-2016-007} the same dataset is used as the one analysed in this paper. 
An attractive alternative is to use model-independent measurements of the strong-phase difference variation over the Dalitz plot, which removes the need to assign model-related systematic uncertainties~\cite{GGSZ,BONDARGGSZ}. 
Measurements of the strong-phase variation in binned regions of the Dalitz plot cannot be done with LHCb data alone, but can be accomplished using an analysis of quantum-correlated neutral $D$ meson pairs from $\psi(3770)$ decays, and have been made at the CLEO-c experiment~\cite{CLEOCISI}. 
These measurements have direct access to the strong-phase difference, which is not the case for the amplitude models based on fits to flavour-tagged \D decays only~\cite{BABAR2008,BABAR2010}. The separation of data into binned regions of the Dalitz plot leads to a loss in statistical sensitivity in comparison to using an amplitude model; however, the advantage of using the measurements from CLEO is that the systematic uncertainties remain free of any model assumptions on the strong-phase difference.  This model-independent method has been used by \belle~\cite{BGGSZ} to study the \BtoDKst, \DtoKsPiPi decay mode, and by \lhcb~\cite{LHCb-PAPER-2014-041} and \belle~\cite{BELLEMODIND} to study \BpmtoDKpm decays.

In this paper, $pp$ collision data at a centre-of-mass energy $\sqrt{s}=7\,(8)\tev$, accumulated by LHCb in 2011 (2012) and corresponding to a total integrated luminosity of $3.0\invfb$, are exploited to perform a model-independent measurement of $\gamma$ in the decay mode \BtoDKst, with \DtoKsPiPi and \DtoKsKK.
The yield of \BtoDKst with \DtoKsPiPi is twice that previously analysed at \belle~\cite{BELLEMODIND} and the \DtoKsKK decay is included for the first time. This allows for a precise measurement of \xy using the techniques developed for similar analyses of $\Bm\to\D\Km$ decays~\cite{LHCb-PAPER-2014-041}.

The remainder of the paper is organised as follows. Section~\ref{sec:principles} describes the analysis framework. Section~\ref{sec:Detector} describes the LHCb detector, and Sect.~\ref{sec:massfit} presents the candidate selection and the parametrisation of the $B$ candidate invariant mass spectrum. 
Section~\ref{sec:dmu} is concerned with the use of semileptonic decays in order to determine the populations in different bins of the $\Dz\to\Kshh$ Dalitz plot. 
Section~\ref{sec:dpfit} discusses the binned Dalitz plot fit and presents the measurements of the \CP violation parameters.  The evaluation of systematic uncertainties is summarised in Sect.~\ref{sec:syst}.  The determination of the CKM angle $\gamma$ using the measured \CP parameters is described in Sect.~\ref{sec:Results}.

\section{Overview of the analysis}
\label{sec:principles}

The favoured and suppressed \Bz decay amplitudes can be expressed as 
\begin{align} 
A(\Bz \to \Dzb X^0_s;p) &\equiv A_c(p)e^{i\delta_c(p)}, \\\nonumber
A(\Bz \to \Dz X^0_s;p) &\equiv A_u(p)e^{i[\delta_u(p)+\gamma]}, 
\end{align}
where $p$ is the $\left(m^2(K\pi), m^2(D\pi)\right)$ coordinate on the $\Bz\to\D K\pi$ Dalitz plot, $A_u(p)$ and $A_c(p)$ are the moduli of the $b \to u$ and $b \to c$ amplitudes, and $\delta_{c,u}(p)$ represent the strong phases of the relevant decay amplitudes. 
The symbol $X^0_s$ refers to a resonant or nonresonant $\Kp\pim$ pair, which could be produced by the decay of the \Kstarz meson or by other contributions to the $\Bz\to\D \Kp\pim$ final state. 
Similar expressions can be written for the \Bzb decay, where the parameter $\gamma$ enters with opposite sign.
The natural width of the \Kstarz (approximately $50\mevcc$~\cite{PDG2014}) must be considered when analysing these decays. In the region near the \Kstarz mass there is interference between the signal \Kstarz decay amplitude and amplitudes due to the other $\Bz\to D\Kp\pim$ Dalitz plot contributions, such as higher mass $K\pi$ resonances and nonresonant $K\pi$ decays. 
Hence, the magnitude ratio between the suppressed and favoured amplitudes $r_{\Bz}$, the coherence factor $\kappa$~\cite{Gronau}, and the effective strong phase difference $\delta_{\Bz}$ depend on the region of the \Bz Dalitz plot to be analysed. These are defined as
\begin{align}
r^2_{\Bz} &\equiv\frac {|A(\Bz \to \Dz \Kstarz)|^2}{|A(\Bz \to \Dzb \Kstarz)|^2}= \frac{ \int_\Kstarz dp \, A_u^2(p)}{ \int_{\Kstarz} dp\, A_c^2(p)}, \\
\kappa e^{i\delta_{\Bz}} &\equiv \frac{  \int_\Kstarz dp\, A_c(p)A_u(p)e^{i[\delta_u(p)-\delta_c(p)]}}{\sqrt{ \int_\Kstarz dp\, A_c^2(p)} \sqrt{ \int_\Kstarz dp\,A_u^2(p)}},
\end{align}
where $0 \leq \kappa \leq 1$. 
For this analysis the integration is over $K^+\pi^-$ masses within $50\mevcc$ of the known \Kstarz mass~\cite{PDG2014} and an absolute value of the cosine of the \Kstarz helicity angle $\theta^\ast$ greater than 0.4. The helicity angle $\theta^\ast$ is defined as the angle between the \Kstarz daughter kaon momentum vector and the direction opposite to the \Bz momentum vector in the \Kstarz rest frame. 
This region is chosen to obtain a large value of $\kappa$ and to facilitate combination with results in Refs.~\cite{LHCb-PAPER-2014-028,LHCb-PAPER-2015-059}, which impose the same limits. 
The coherence factor has recently been determined by the LHCb collaboration to be $\kappa = 0.958\,^{+0.005}_{-0.010}\,^{+0.002}_{-0.045}$~\cite{LHCb-PAPER-2015-059}, through an amplitude analysis that measures the $b\to c$ and $b \to u$ amplitudes in the $\Bz \to D K^+ \pi^-$ decay.

The amplitude of the \Dz meson decay at a particular point on the $D$ Dalitz plot is defined as $A_D(\msqmin,\msqplus) \equiv A(\msqmin,\msqplus)e^{i\delta(\msqmin,\msqplus)}$, where \msqmin (\msqplus) is the invariant mass of the $\KS h^-$ ($\KS h^+$) pair. Neglecting \CP violation in charm decays, which is known to be small~\cite{PDG2014}, the charge-conjugated amplitudes are related by $A_{\overline{D}}(\msqmin,\msqplus) = A_D(\msqplus,\msqmin)$. The partial widths for the $B$ decays can be written as 
\begin{align}
d{\Gamma}  (\Bzb \to D(\to\Kshh) &\overline{X}^0_s; p,\msqmin,\msqplus) \propto  \\ \nonumber
	& \bigl\lvert A_c(p)e^{i\delta_c(p)}A_D(\msqmin,\msqplus) + A_u(p)e^{i[\delta_u(p)-\gamma]}A_{\overline{D}}(\msqmin,\msqplus)\bigr\rvert^2, \\
d{\Gamma}  (\Bz \to D(\to\Kshh)& {X^0_s};  p,\msqmin,\msqplus) \propto \\ \nonumber
  & \bigl\lvert A_c(p)e^{i\delta_c(p)}A_{\overline{D}}(\msqmin,\msqplus) + A_u(p)e^{i[\delta_u(p)+\gamma]}A_D(\msqmin,\msqplus) \bigr\rvert^2.
\end{align}
Expanding and integrating over the defined \Kstarz region, one obtains
\begin{align}
\label{eq:dp}
d{\Gamma}& (\Bzb \to D(\to\Kshh) \Kstarzb; \msqmin,\msqplus) \propto \\ \nonumber
& \bigl\lvert A_D(\msqmin,\msqplus)\bigr\rvert ^2 + r_{\Bz}^2 \bigl\lvert A_D(\msqplus,\msqmin)\bigr\rvert ^2 + 2\kappa r_{\Bz} \mathrm{Re}\big[A_D(\msqmin,\msqplus)A^*_D(\msqplus,\msqmin)e^{-i(\delta_{\Bz}-\gamma)}\big], \\
\label{eq:dp2}
d{\Gamma}& (\Bz \to D(\to\Kshh) {\Kstarz};\msqmin,\msqplus) \propto \\ \nonumber
& \bigl\lvert A_D(\msqplus,\msqmin)\bigr\rvert ^2 + r_{\Bz}^2 \bigl\lvert A_D(\msqmin,\msqplus)\bigr\rvert ^2 + 2\kappa r_{\Bz} \mathrm{Re}\big[A_D(\msqplus,\msqmin)A^*_D(\msqmin,\msqplus)e^{-i(\delta_{\Bz}+\gamma)}\big]. 
\end{align}

The $D$ Dalitz plot is partitioned into bins symmetric under the exchange $\msqmin \leftrightarrow \msqplus$. The cosine of the strong-phase difference between the \Dz and \Dzb decay weighted by the decay amplitude and averaged in bin $i$ is called $c_i$~\cite{GGSZ,BONDARGGSZ}, and is given by 
\begin{align}
c_i &\equiv \frac{\int_{i} d\msqmin \, d\msqplus \, A(\msqmin,\msqplus) A(\msqplus,\msqmin) \cos[\delta(\msqmin,\msqplus)-\delta(\msqplus,\msqmin)]}
{\sqrt{\int_{i} d\msqmin \, d\msqplus \, A^2(\msqmin,\msqplus) \int_{i} d\msqmin \, d\msqplus \, A^2(\msqplus,\msqmin)}},
\label{eq:ci}
\end{align}
where the integrals are evaluated over the phase space of bin $i$. An analogous expression can be written for $s_i$ which is the sine of the strong-phase difference weighted by the decay amplitude and averaged in the bin.

\begin{figure}[tb]
\centering
\includegraphics[width=0.48\textwidth]{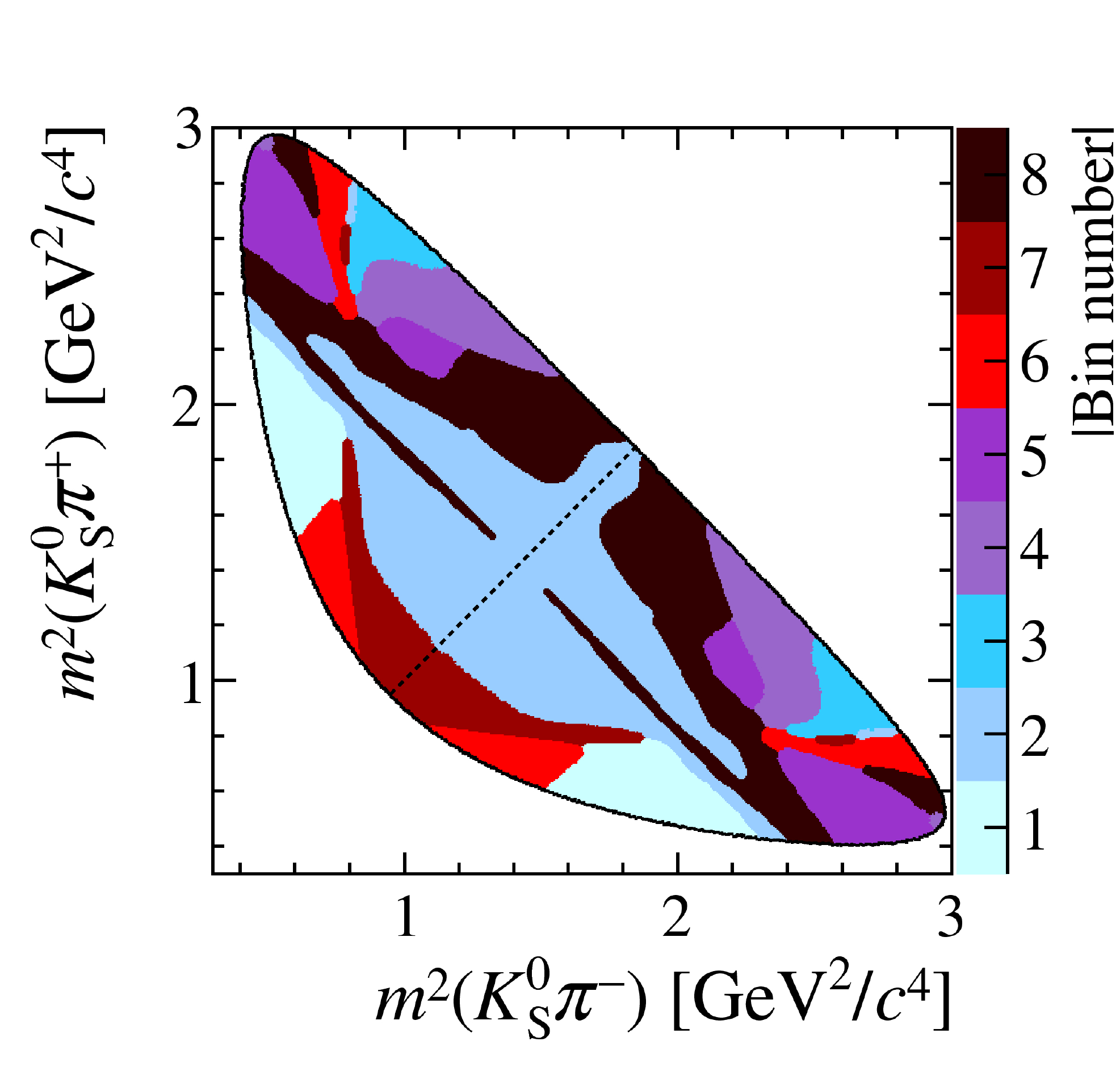}
\hfill
\includegraphics[width=0.48\textwidth]{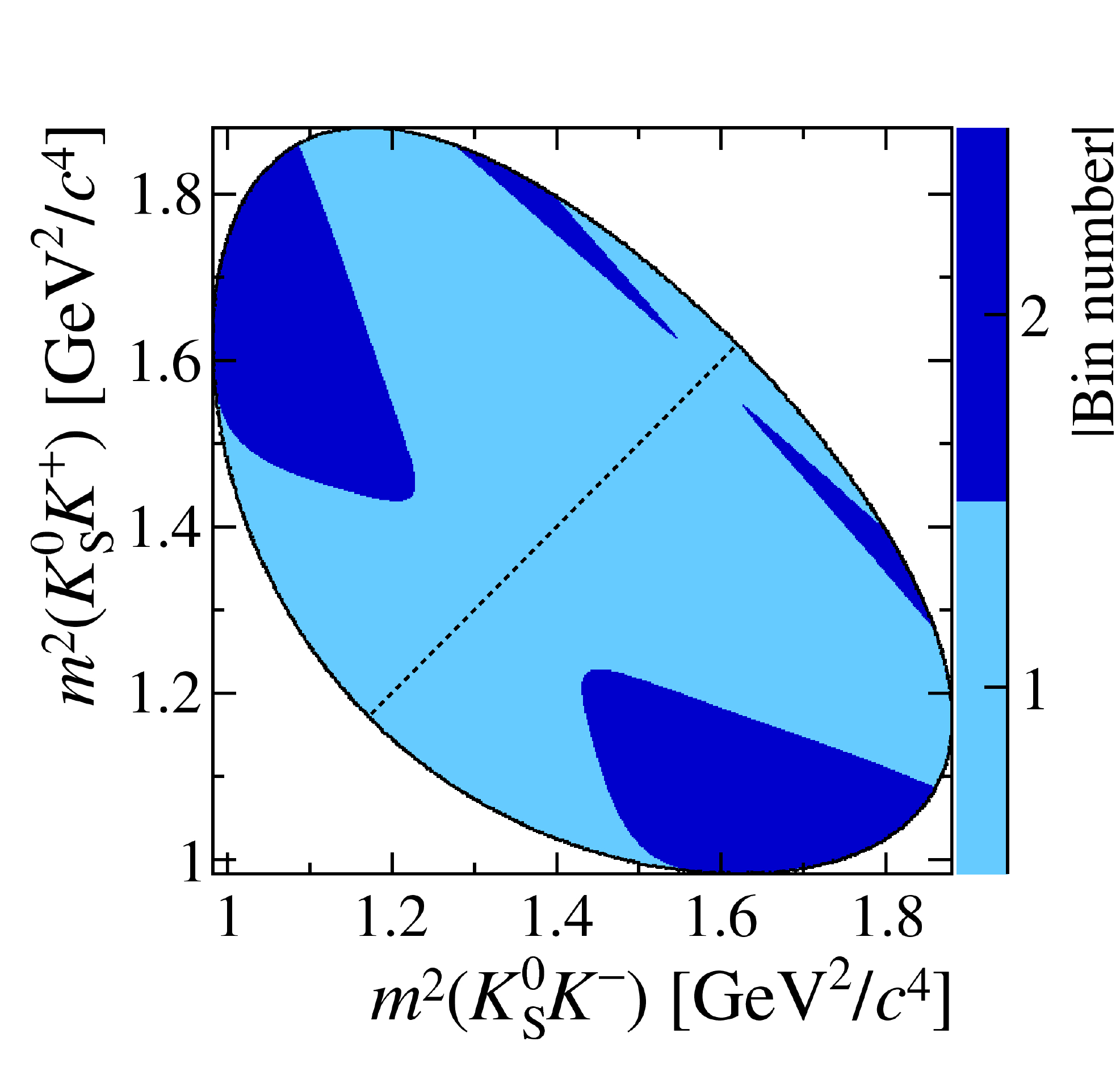}
\caption{Binning schemes for (left) \DtoKsPiPi and (right) \DtoKsKK. The diagonal line separates the positive and negative bin numbers, where the positive bins are in the region $m^2_- \geq m^2_+$.}
\label{fig:bins}
\end{figure}

Measurements of $c_i$ and $s_i$ are provided by CLEO in four different $2\times 8$ binning schemes for the \DtoKsPiPi decay~\cite{CLEOCISI}. The bins are labelled from $-8$ to $+8$, excluding zero, where the bins containing a positive label satisfy the condition $\msqmin \ge \msqplus$. The binning scheme used in this analysis is referred to as the `modified optimal' binning. The optimisation was performed assuming a strong-phase difference distribution given by the \babar model presented in Ref.~\cite{BABAR2008}. This modified optimal binning is described in Ref.~\cite{CLEOCISI} and was designed to be statistically optimal in a scenario where the signal purity is low. It is also more robust for analyses with low yields in comparison to the alternatives, as no individual bin is very small. 
For the \KsKK final state, the measurements of $c_i$ and $s_i$ are available in three variants containing a different number of bins, with the guiding model being that from the \babar study described in Ref.~\cite{BABAR2010}. For the present analysis the variant with the $2\times 2$ binning is chosen, given the very low signal yields expected in this decay.
The measurements of $c_i$ and $s_i$ are not biased by the use of a specific amplitude model in defining the bin boundaries, which only affects this analysis to the extent that if the model gives a poor description of the underlying decay then there will be a reduction in the statistical sensitivity of the $\gamma$ measurement. The binning choices for the two decay modes are shown in Fig.~\ref{fig:bins}.

The integrals of Eqs.~(\ref{eq:dp}) and (\ref{eq:dp2}) over the phase space of a Dalitz plot bin are proportional to the expected yield in that bin. 
The physics parameters of interest, $r_{\Bz}$, $\delta_{\Bz}$, and $\gamma$, are translated into four Cartesian variables~\cite{BABAR2005,Yabsley}. These are the measured observables and are defined as
\begin{equation}
x_\pm \equiv r_{\Bz} \cos (\delta_{\Bz} \pm \gamma)\; \mathrm{and} \; y_\pm \equiv r_{\Bz} \sin (\delta_{\Bz} \pm \gamma).
\label{eq:xydefinitions}
\end{equation}
From Eqs.~(\ref{eq:dp}) and (\ref{eq:dp2}) it follows that
\begin{align}
\label{eq:populations}
N_{\pm i}^+ &= n_{+} \left[ F_{\mp i} + (x_+^2 + y_+^2) F_{\pm i} + 2 \kappa\sqrt{F_{+i} F_{-i}} ( x_+ c_{\pm i} - y_+ s_{\pm i}) \right], \\
\label{eq:populations2}
N_{\pm i}^- &= n_{-} \left[ F_{\pm i} + (x_-^2 + y_-^2) F_{\mp i} + 2 \kappa\sqrt{F_{+i} F_{-i}} ( x_- c_{\pm i} + y_- s_{\pm i}) \right],
\end{align}
where $F_i$ are defined later in Eq.~(\ref{eq:fi}) and $N^{+}_i$ ($N^-_i$) is the expected number of \Bz (\Bzb) decays in bin $i$. The superscript on $N$ refers to the charge of the kaon from the \Kstarz decay. The parameters $n_+$ and $n_-$ provide the normalisation, which can be different due to production, detection and \CP asymmetries between \Bz and \Bzb mesons. 
However the integrated yields are not used and the analysis is insensitive to such effects.
The detector and selection requirements placed on the data lead to a non-uniform efficiency over the Dalitz plot. The efficiency profile for the signal candidates is given by $\eta=\eta(\msqmin,\msqplus)$. Only the relative efficiency from one point to another matters and not the absolute normalisation. 
The parameters $F_i$ are given by
\begin{equation}
\label{eq:fi}
F_i = \frac{\int_{i} d\msqmin d\msqplus |A_{D}(\msqmin,\msqplus)|^2 \, \eta(\msqmin,\msqplus) }{\sum_j \int_{j} d\msqmin d\msqplus |A_{D}(\msqmin,\msqplus)|^2\, \eta(\msqmin,\msqplus) }
\end{equation}
and are the fraction of decays in bin $i$ of the $\Dz\to\Kshh$ Dalitz plot. 
 
The values of $F_i$ are determined from the control decay mode \BztoDstmuX, where the \Dstarm decays to \Dzb\pim and the \Dzb decays to either the \KsPiPi or \KsKK final state. The symbol $X$, hereinafter omitted, indicates other particles which may be produced in the decay but are not reconstructed. Samples of simulated events are used to correct for the small differences in efficiency arising through necessary differences in selecting \BztoDstmu and \BtoDKst decays, which are discussed further in Sect.~\ref{sec:dmu}. 

Effects due to \Dz-\Dzb mixing and \CP violation in \Kz-\Kzb mixing are ignored: the corrections are discussed in Refs.~\cite{BPV,Yuval} and are expected to be of order $0.2\degrees$ ($1\degrees$) for $D$ mixing (\CP violation in $K$ mixing) in \BmtoDKm decays. In both cases the size of the correction is reduced as the value of $r_{\Bz}$ is expected to be approximately three times larger than the value of $r_B$ in \BmtoDKm decays. 
The effect of different nuclear interactions within the detector material for \Kz and \Kzb mesons is expected to be of a similar magnitude and is also ignored~\cite{LHCb-PAPER-2014-013}.

\section{Detector and simulation}
\label{sec:Detector}
The \lhcb detector~\cite{Alves:2008zz,LHCb-DP-2014-002} is a single-arm forward
spectrometer covering the \mbox{pseudorapidity} range $2<\eta <5$,
designed for the study of particles containing \bquark or \cquark
quarks. The detector includes a high-precision tracking system
consisting of a silicon-strip vertex detector surrounding the $pp$
interaction region, a large-area silicon-strip detector located
upstream of a dipole magnet with a bending power of about
$4{\mathrm{\,Tm}}$, and three stations of silicon-strip detectors and straw
drift tubes placed downstream of the magnet.
The tracking system provides a measurement of momentum, \ptot, of charged 
particles with a relative uncertainty that varies from 0.5\% at low momentum to 
1.0\% at 200\gevc.
The minimum distance of a track to a primary vertex (PV), the impact parameter 
(IP), is measured with a resolution of $(15+29/\pt)\mum$,
where \pt is the component of the momentum transverse to the beam, in\,\gevc.
Different types of charged hadrons are distinguished using information
from two ring-imaging Cherenkov detectors. 
Photons, electrons and hadrons are identified by a calorimeter system consisting of
scintillating-pad and preshower detectors, an electromagnetic
calorimeter and a hadronic calorimeter. Muons are identified by a
system composed of alternating layers of iron and multiwire
proportional chambers. 
The online event selection is performed by a trigger, which consists of a hardware 
stage, based on information from the calorimeter and muon systems, followed by a 
software stage, which applies a full event reconstruction. 
The trigger algorithms used to select hadronic and semileptonic $B$ decay 
candidates are slightly different, due to the presence of the muon in the latter, 
and are described in Sects.~\ref{sec:massfit} and \ref{sec:dmu}.

In the simulation, $pp$ collisions are generated using 
\pythia~\cite{Sjostrand:2006za,Sjostrand:2007gs} 
with a specific \lhcb configuration~\cite{LHCb-PROC-2010-056}. Decays of hadronic 
particles are described by \evtgen~\cite{Lange:2001uf}, in which final-state radiation 
is generated using \photos~\cite{Golonka:2005pn}. The interaction of the generated 
particles with the detector, and its response, are implemented using the \geant 
toolkit~\cite{Allison:2006ve, Agostinelli:2002hh} as described in 
Ref.~\cite{LHCb-PROC-2011-006}.

\section{\boldmath Event selection and fit to the \texorpdfstring{$B$}{B} candidate invariant mass distribution}
\label{sec:massfit}

Decays of the \KS meson to the $\pip\pim$ final state are reconstructed in two different categories, the first involving \KS mesons that decay early enough for the pion track segments to be reconstructed in the vertex detector, the second containing \KS mesons that decay later such that track segments of the pions cannot be formed in the vertex detector. These categories are referred to as \emph{long} and \emph{downstream}. The candidates in the long category have better mass, momentum, and vertex resolution than those in the downstream category. 

Signal events considered in the analysis must first fulfil hardware and software trigger requirements. At the hardware stage at least one of the two following criteria must be satisfied: either a particle produced in the decay of the signal $B$ candidate leaves a deposit with high transverse energy in the hadronic calorimeter, or the event is accepted because particles not associated with the signal candidate fulfil the trigger requirements.
At least one charged particle should have a high \pt and a large \chisqip with respect to any PV, where \chisqip is defined as the difference in \chisq of a given PV fitted with and without the considered track.
At the software stage, a multivariate algorithm~\cite{BBDT} is used for the identification of secondary vertices that are consistent with the decay of a \bquark hadron.
The software trigger designed to select \BtoDKst candidates requires a two-, three- or four-track secondary vertex with a large scalar sum of the \pt of the associated charged particles and a significant displacement from the PVs.
The PVs are fitted with and without the \B candidate tracks, and the PV that gives the smallest \chisqip is associated with the \B candidate.

Combinatorial background is rejected primarily through the use of a multivariate approach with a boosted decision tree (BDT)~\cite{Breiman,AdaBoost}. 
The signal and background training samples for the BDT are simulated signal events and candidates in data with reconstructed $B$ candidate mass in a sideband region.
Loose selection criteria are applied to the training samples on all intermediate states ($D$, \KS, \Kstarz).
Separate BDTs are trained for candidates containing long and downstream \KS candidates. 
Due to the presence of the topologically indistinguishable \BstoDKst decay, the available background event sample for the training is limited to the mass range 5500--$6000\mevcc$.  
To make full use of all background candidates for the training of the BDTs, all events are divided into two sets at random. For each \KS category two BDTs are trained, using each set of events in the sideband. The results of each BDT training are applied to the events in the other sample. Hence, in total four BDTs are trained, and in this way the BDT applied to one set of events is trained with a statistically independent set of events.

Each BDT uses a total of 16 variables, of which the most discriminating are the \chisq of the kinematic fit of the whole decay chain (described below), the \Kstarz transverse momentum, and the flight distance significance of the $B$ candidate from the associated PV. 
In the BDT for long \KS candidates, two further variables are found to provide high separation power: the flight distance significance of the \KS decay vertex from the PV and 
a variable characterising the flight distance significance of the \KS vertex from the $D$ vertex along the beam line.
The remaining variables in the BDT are 
the \chisqip of the $B$ candidate, 
the sum of \chisqip of the two \KS daughter tracks, 
the sum of the \chisqip of all the other tracks, 
the vertex quality of the $B$ and $D$ candidates, 
the flight distance significance of the $D$ vertex from the PV,
a variable characterising the flight distance significance between the $D$ and $B$ vertices along the beam line, 
the transverse momentum of each of the $D$ and $B$ candidates, 
the cosine of the angle between the $B$ momentum vector and the vector between the production and decay vertex,
and the helicity angle $\theta^\ast$.
It has been verified that the use of $\theta^\ast$ in the BDT has no significant impact on the value of $\kappa$. 
An optimal criterion on the BDT discriminator is determined with a series of pseudoexperiments to obtain the value that provides the best sensitivity to \xy. 

A kinematic fit~\cite{Hulsbergen:2005pu} is imposed on the full $B$ decay chain. The fit constrains the $B$ candidate to point towards the PV, and the \D and \KS candidates to have their known masses~\cite{PDG2014}. 
This fit improves the $B$ mass resolution and therefore provides greater discrimination between signal and background; furthermore, it improves the resolution on the Dalitz plot and ensures that all candidates lie within the kinematically-allowed region of the \DtoKshh Dalitz plot.
The kinematic variables obtained in this fit are used to determine the physics parameters of interest and the \chisq of this fit is used in the BDT training.
 
To suppress background further, particle identification (PID) requirements are placed on both daughter tracks of the \Kstarz to identify the kaon and the pion. This also removes the possibility of a second \Kstarz candidate being built from the same pair of tracks with opposite particle hypotheses. The PID requirement on the kaon is tight, with an efficiency of 81\%, and is necessary to suppress 98\% of the background from \BtoDrho decays where a pion from the \rhoz decay is misidentified as a kaon. The absolute value of $\cos\theta^\ast$ is required to be greater than 0.4, as discussed in Sect.~\ref{sec:principles}. 

For the selection on the $D$ (\KS) mass, the mass is computed from a kinematic fit~\cite{Hulsbergen:2005pu} that constrains the \KS (\D) mass to its known value and the \B candidate to point towards the PV.
The $D$ meson mass is required to be within $30\mevcc$ of the nominal mass~\cite{PDG2014} which is three times the mass resolution. The long (downstream) \KS candidates are required to be within $14.4\,(19.9)\mevcc$ of their nominal mass which again corresponds to three times the mass resolution.  In the case of \DtoKsKK candidates a loose PID cut is also placed on the kaon daughters of the $D$ to remove cross-feed from other $D \to \Kshh$ decays. One further physics background is due to $D$ decays to four pions where two pions are consistent with a long \KS candidate. To suppress this background to negligible levels, a tight requirement is placed on the flight distance significance of the long \KS candidate from the $D$ vertex along the beam line.

While the selection is different for long and downstream \KS candidates, the small differences between the \B candidate mass resolution for the two categories observed in simulation are negligible for this analysis.
This is because of the $D$ mass constraint applied in the kinematic fit. Therefore, both \KS categories are combined in the fit of the $B$ invariant mass distribution. All $B$ meson candidates with invariant mass between $5200$ and $5800\mevcc$ are fitted together to obtain the signal and background yields. 

\begin{figure}[tb]
\centering
\includegraphics[width=0.69\textwidth]{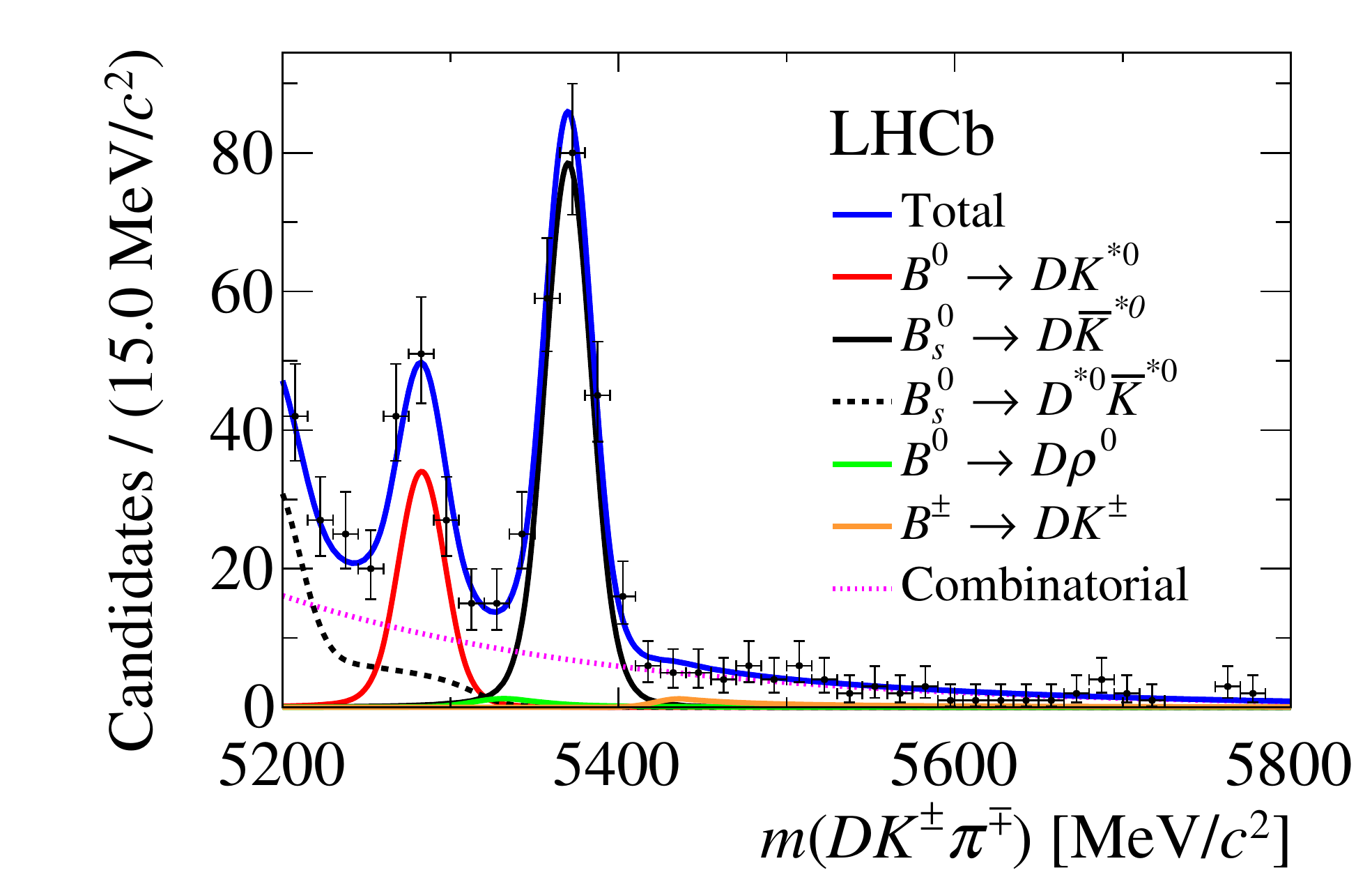} \\
\includegraphics[width=0.69\textwidth]{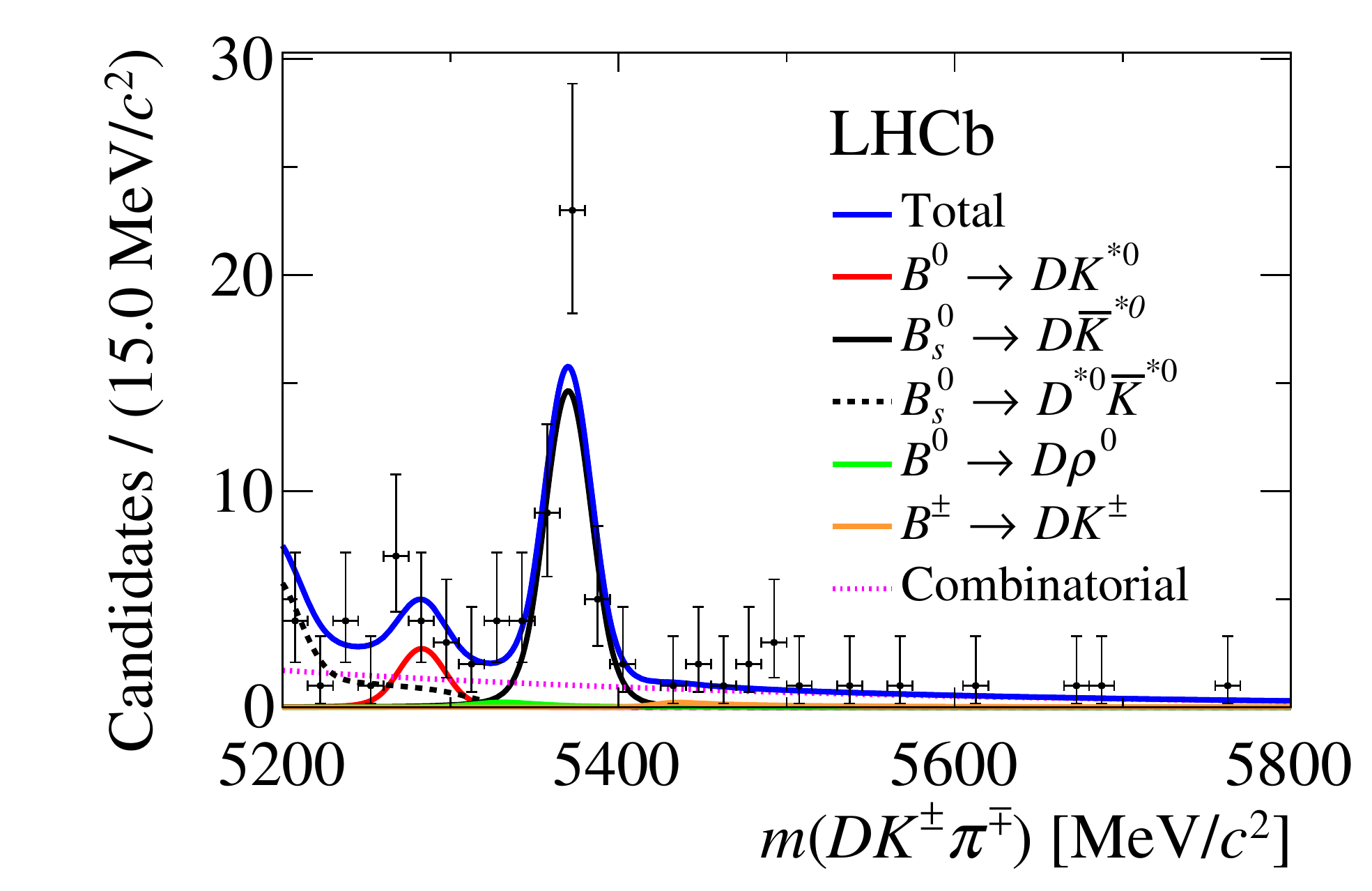}
\caption{Invariant mass distributions of \BtoDKst candidates with (top) \DtoKsPiPi and (bottom) \DtoKsKK. The fit results, including the signal and background components, are superimposed.}
\label{fig:mass_kspipi}
\end{figure}

The invariant mass distributions of the selected candidates are shown in Fig.~\ref{fig:mass_kspipi} for both \D decay modes. The \Bz and \Bzb candidates are summed.
The result of an extended maximum likelihood fit to these distributions is superimposed.
The fit is performed simultaneously for candidates from both $D$ decays, allowing parameters, unless otherwise stated, to be common between both $D$ decay categories.
Figure~\ref{fig:mass_kspipi} shows the various components that are considered in the fit to the invariant mass spectra. In addition to the signal \BtoDKst component, there are contributions from \BstoDKst, from \BtoDrho where one pion is misidentified as a kaon, and from \BtoDK decays where one pion from the rest of the event is added to create a fake \Kstarz. A large background comes from \BstoDstKst decays where the photon or neutral pion from the $\Dstarz$ decay is not reconstructed.
The purpose of this fit is to determine the parametrisation of the signal and background components, and the size of the background contributions, which are used in the fit of partitioned regions of the Dalitz plot described in Sect.~\ref{sec:dpfit}.

The \BstoDKst and \BtoDKst decays are modelled by the same probability density function (PDF), a sum of two Crystal Ball~\cite{Skwarnicki:1986xj} functions with common mean and width parameters. The mean for the \Bs meson is determined in the fit and the mean for the \Bz meson is required to be $87.19\mevcc$~\cite{PDG2014} lower. The width is allowed to vary in the fit and is required to be the same for the two decays. All other parameters are fixed from simulation. The combinatorial background is modelled by an exponential function with slope determined by the fit for the \DtoKsPiPi and \DtoKsKK categories separately. The PDF for \BtoDrho decays is derived from simulation with additional data-driven corrections applied to take into account PID response differences between data and simulation~\cite{LHCb-PROC-2011-008}. This background is described with the sum of two Crystal Ball functions, whose parameters are obtained from the weighted simulated events. The \BtoDK background is treated in a similar fashion. 

For the partially reconstructed background from \BstoDstKst decays the distribution in the invariant mass spectrum is dependent on the helicity state of the \Dstarz meson. The initial decay of the \Bs involves the decay of a pseudoscalar to two vector particles. Hence, due to angular momentum conservation there are three helicity amplitudes to consider, which can be labelled by the \Dstarz helicity state $\lambda = -1, 0, +1$. 
In the subsequent parity-conserving decay $\Dstarz \to \Dz \{\piz,\gamma\}$, the value of $\lambda$ and the spin of the missing neutral particle determines the distribution of the \Dstarz helicity angle, which is defined as the angle between the missing neutral particle's momentum vector and the direction opposite to the \B meson in the \Dstarz rest frame. 
The resulting distributions for $\lambda= -1$ or $+1$ are identical and hence are grouped together.
The functional forms of the underlying $DK\pi$ invariant mass spectrum, shown in Table~\ref{tab:iniPDFs}, can be calculated based on $\lambda$, and the spin and mass of the missing particle. The parameters $a_X$ and $b_X$ are the kinematic endpoints of the reconstructed $DK\pi$ invariant mass, where $X$ is the particle that is not reconstructed. 
These distributions are further modified to take into account detector resolution and reconstruction efficiency. The parameters for the resolution and efficiency are determined from fits to simulated samples, while the endpoints are calculated using the masses of the particles involved. 

\begingroup 
\renewcommand{\arraystretch}{1.7}
\begin{table}[tbp]
\caption{Functional forms of the $DK\pi$ invariant mass distribution, $m$, in partially reconstructed decays of $\Bs\to(\Dstarz\to\Dz\{\piz,\gamma\})\Kstarzb$, where either the \piz or $\gamma$ is not reconstructed. The \Dstarz helicity state is given by $\lambda$. The quantities $a_X$ and $b_X$ are the minimum and maximum kinematic boundaries of the reconstructed $DK\pi$ invariant mass, where $X$ is the particle that is missed.}
\label{tab:iniPDFs}
\centering
\begin{tabular}{c c@{\hskip 20pt} c}
Missed particle & $\lambda$ & PDF \\
\hline
$\piz$ & 0 & $\left(m - \frac{a_\piz+b_\piz}{2} \right)^2$ \\
$\piz$ & $-1$ or $+1$ & $- (m-a_\piz)(m-b_\piz)$ \\
$\gamma$ & 0 & $- (m-a_\gamma)(m-b_\gamma)$ \\
$\gamma$ & $-1$ or $+1$ & $\left(m - \frac{a_\gamma+b_\gamma}{2} \right)^2 + \left (\frac{a_\gamma-b_\gamma}{2} \right)^2$\\
\end{tabular}
\end{table}
\endgroup

The lower range of the mass fit is $5200\mevcc$. The removal of candidates with invariant mass below this value reduces the background from \BtoDstKst decays to a small level, which is neglected in the baseline fit. Other contributions such as $B^\pm \to D h^\pm \pip \pim$, where one particle is missing and another may be misidentified, are also reduced to a negligible level. 

With the large number of overlapping signal and background contributions it is not possible to let all yield parameters vary freely, especially as some background contributions are expected to have small yields. Therefore, the strategy employed is to constrain the ratio of these background yields to the \BstoDKst contribution. The constraints are determined by taking into account all relevant branching fractions~\cite{PDG2014}, fragmentation fractions~\cite{HFAG} and selection efficiencies determined from simulation. This is possible for the contributions \BtoDrho and \BtoDK where the branching fractions are measured. The ratio of \BtoDrho (\BtoDK) to \BstoDKst is constrained in the fit to $R_{\rho}=(2.9\pm0.8)\%$ ($R_{DK}=(4.2\pm1.0)\%$). In the case of the \BstoDstKst background, neither its branching fraction nor the relative fraction of the $\Dstarz$ helicity states has been measured. Therefore, information is taken from the higher statistics $\Bs\to D(\to K\pi)\Kstarzb$ decay, which has been studied by the LHCb collaboration~\cite{LHCb-PAPER-2014-028}.
In these Cabibbo-favoured decays the mass distribution is simpler since the \BtoDKst and \BtoDstKst decays are doubly Cabibbo-suppressed, hence allowing the shape parameters and yields for the \BstoDKst and \BstoDstKst decays to be reliably determined. 
The expected ratio $R_s$ between \BstoDstKst and \BstoDKst can be determined using the information from the analysis of $D\to K\pi$ decays, with a correction for the selection efficiencies.
The ratio between the total yield of the \BstoDstKst candidates with reconstructed mass above $5200\mevcc$ and \BstoDKst candidates is determined to be $R_s=(35\pm 14)\%$. The fraction of \BstoDstKst candidates where $\lambda=0$ is determined to be $\alpha=0.72\pm0.13$.
The yields of the \BstoDKst, \BtoDKst and the combinatorial background are free parameters in the fit. Pseudoexperiments for this fit configuration show that only negligible biases are expected. The fitted yields and parameters of the fit are given in Table~\ref{tab:massfit:results}. The purity in the signal region, defined as $\pm 25\mevcc$ around the \Bz mass measured in the fit, is 59\% (44\%) for the \KsPiPi  (\KsKK) candidates. The background is dominated by combinatorial and \BstoDstKst decays. Contributions from the other backgrounds considered are small.

\begingroup 
\renewcommand{\arraystretch}{1.2}
\begin{table}[tbp]
\begin{center}
\caption{Results of the simultaneous fit to the invariant mass distribution of \BtoDKst decays, with the \D meson decaying to \KsPiPi and \KsKK.}
\label{tab:massfit:results}
\begin{tabular}{l r@{\hskip 1pt} l}
Variable & \multicolumn{2}{l}{Fitted value and uncertainty} \\
\hline
\Bs mass                    & $5369.2$ & $^{+1.0}_{-1.0}\;\mevcc$ \\
Signal width parameter      & $13.3$ & $^{+1.0}_{-0.9}\;\mevcc$ \\
\KsKK exponential slope     & $(-3.4$ & $^{+1.6}_{-1.4})\e{-3}\;(\!\mevcc)^{-1}$ \\
\KsPiPi exponential slope   & $(-5.4$ & $^{+0.9}_{-0.8})\e{-3}\;(\!\mevcc)^{-1}$ \\
$\alpha$                    & $0.74$ & $^{+0.13}_{-0.13}$ \\
$R_{DK}$                    & $(4.3$ & $^{+1.0}_{-1.0})\e{-2}$ \\
$R_\rho$                    & $(3.0$ & $^{+0.8}_{-0.8})\e{-2}$ \\
$R_s$                       & $0.31$ & $^{+0.09}_{-0.09}$ \\
$n(\BtoDKst, \KsPiPi)$      & $84$ & $^{+15}_{-14}$ \\
$n(\BstoDKst, \KsPiPi)$     & $194$ & $^{+18}_{-17}$ \\
$n(\mathrm{combinatorial}, \KsPiPi)$ & $207$ & $^{+36}_{-35}$ \\
$n(\BtoDKst, \KsKK)$        & $6.7$ & $^{+4.8}_{-4.2}$ \\
$n(\BstoDKst, \KsKK)$       & $36.3$ & $^{+7.1}_{-6.4}$ \\
$n(\mathrm{combinatorial}, \KsKK)$   & $32.3$ & $^{+10.0}_{-9.0}$ \\
\end{tabular}
\end{center}
\end{table}
\endgroup

\begin{figure}[tb]
\centering
\includegraphics[width=0.45\textwidth]{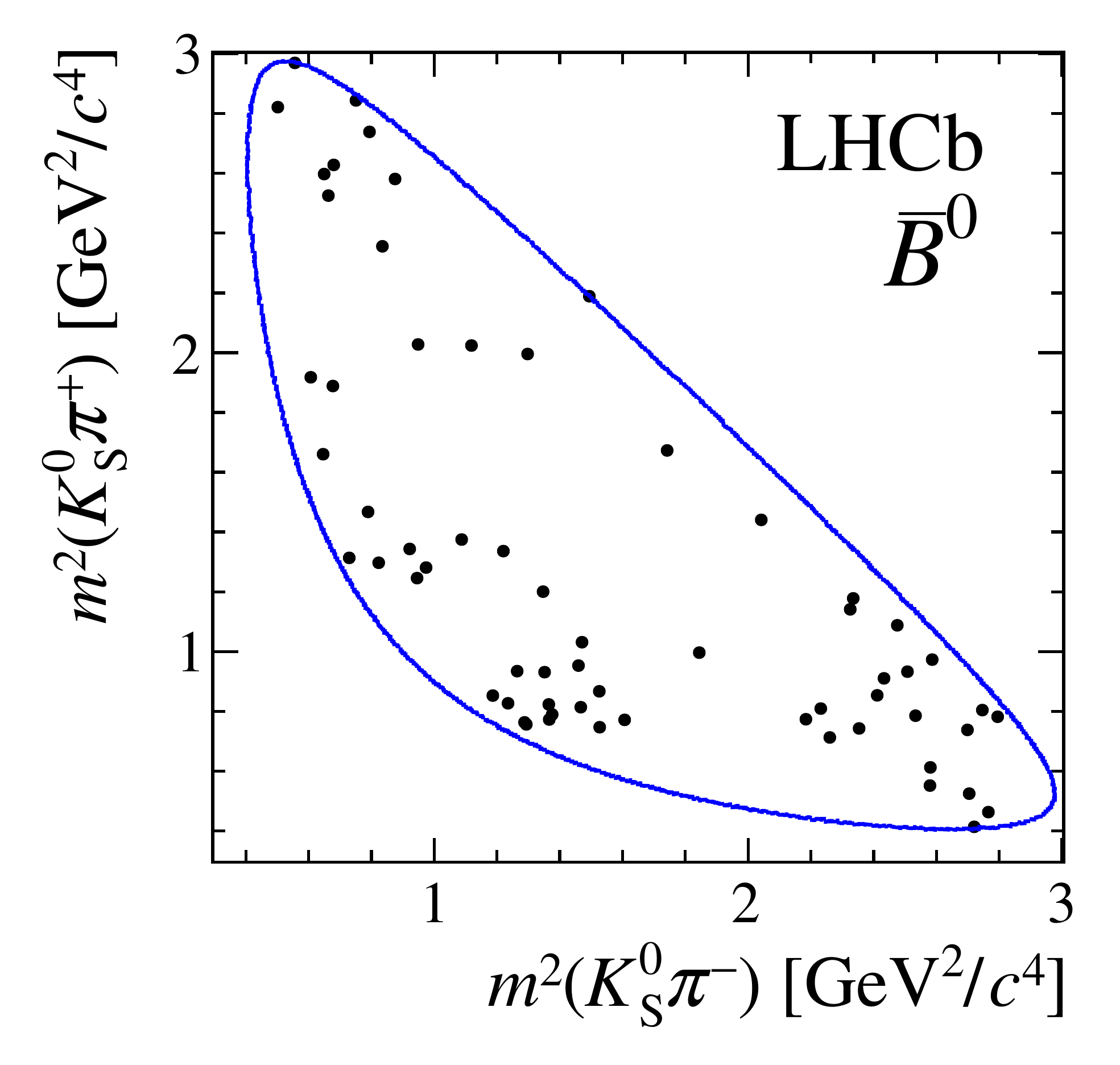}
\includegraphics[width=0.45\textwidth]{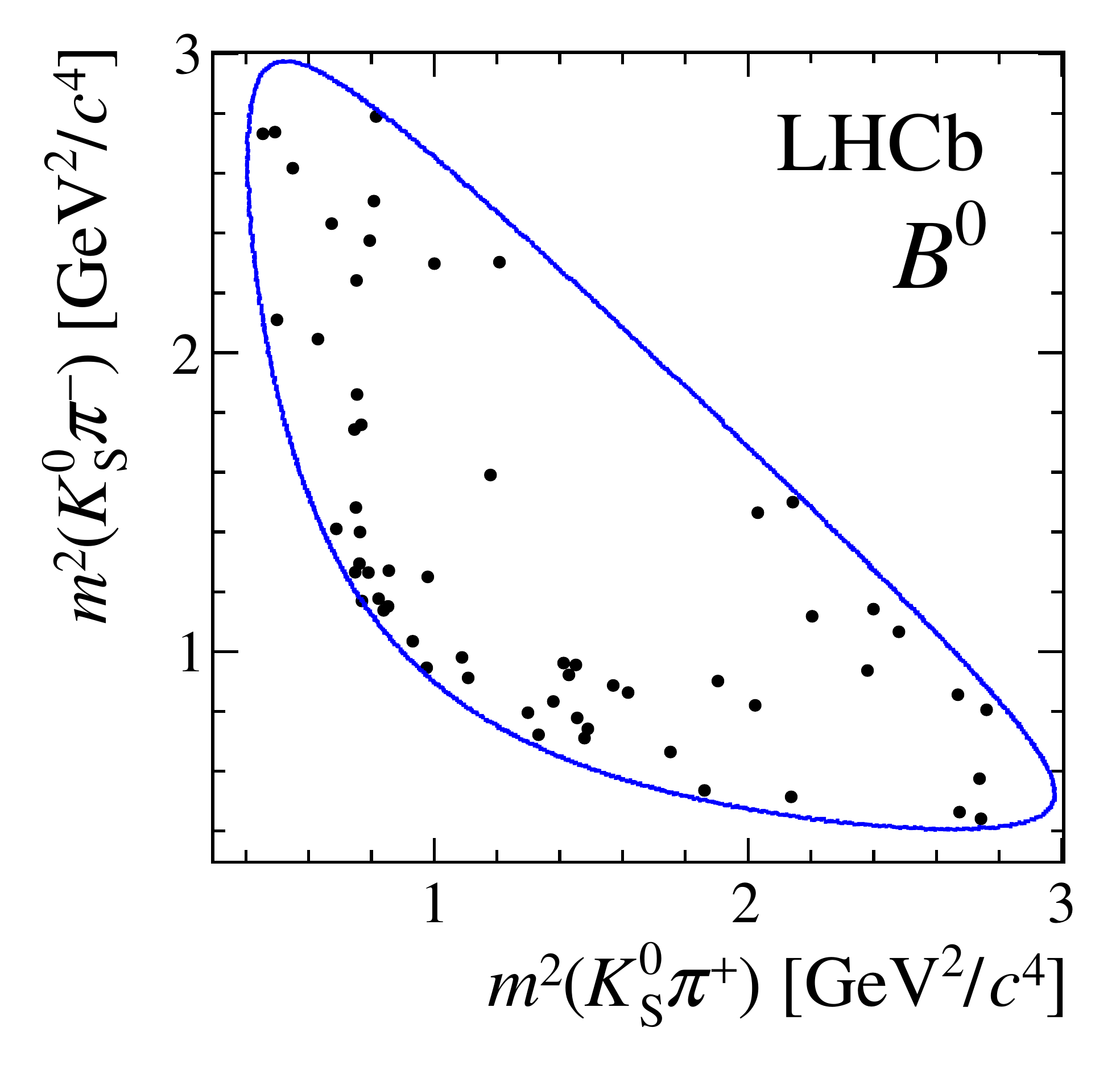}
\caption{Dalitz plots of candidates in the signal region for \DtoKsPiPi decays from (left) $\Bzb\to D\Kstarzb$ and (right) $\Bz\to D\Kstarz$ decays. The solid blue line indicates the kinematic boundary.}
\label{fig:dalitzpipi}
\end{figure}

\begin{figure}[tb]
\centering
\includegraphics[width=0.45\textwidth]{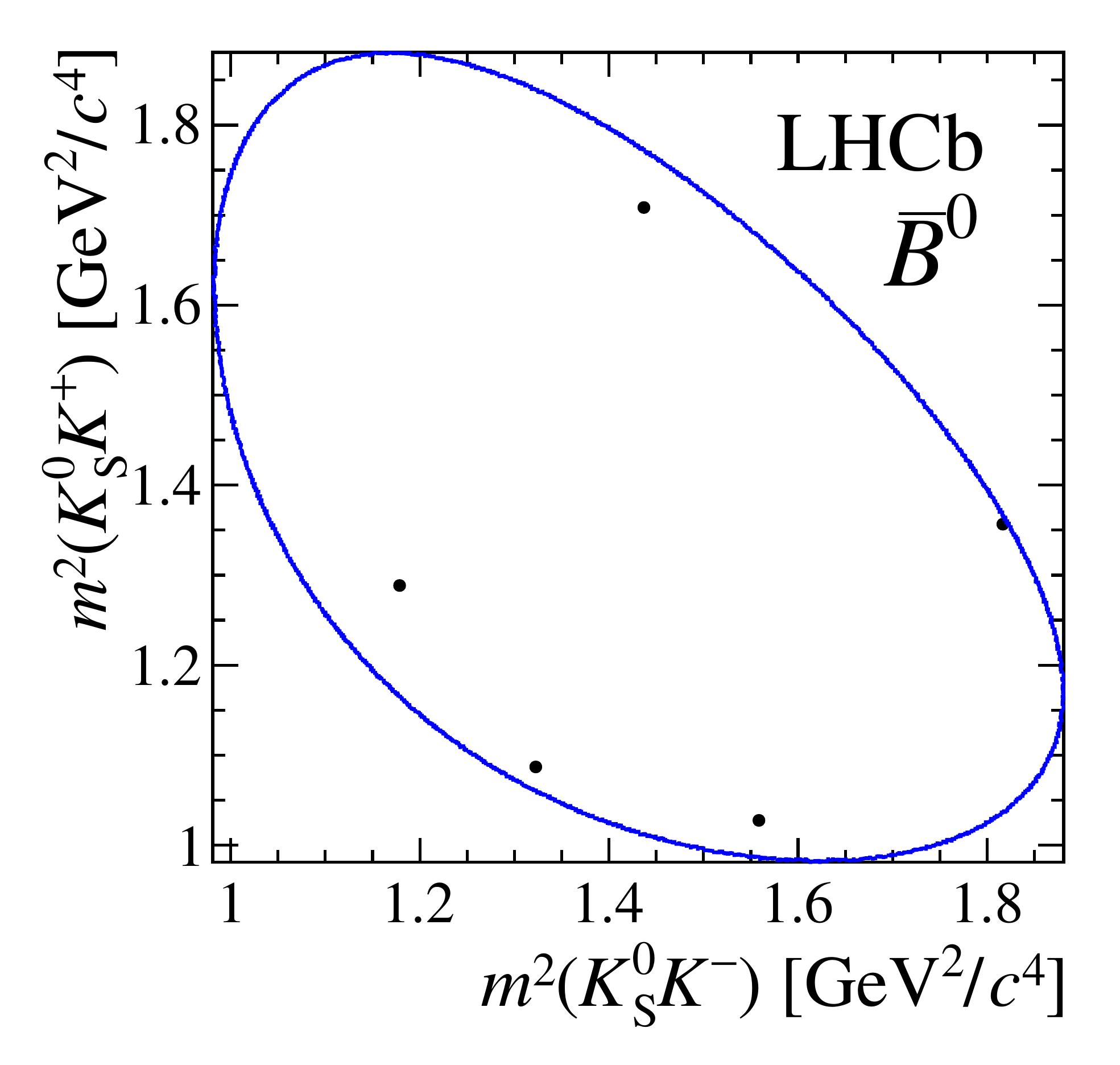}
\includegraphics[width=0.45\textwidth]{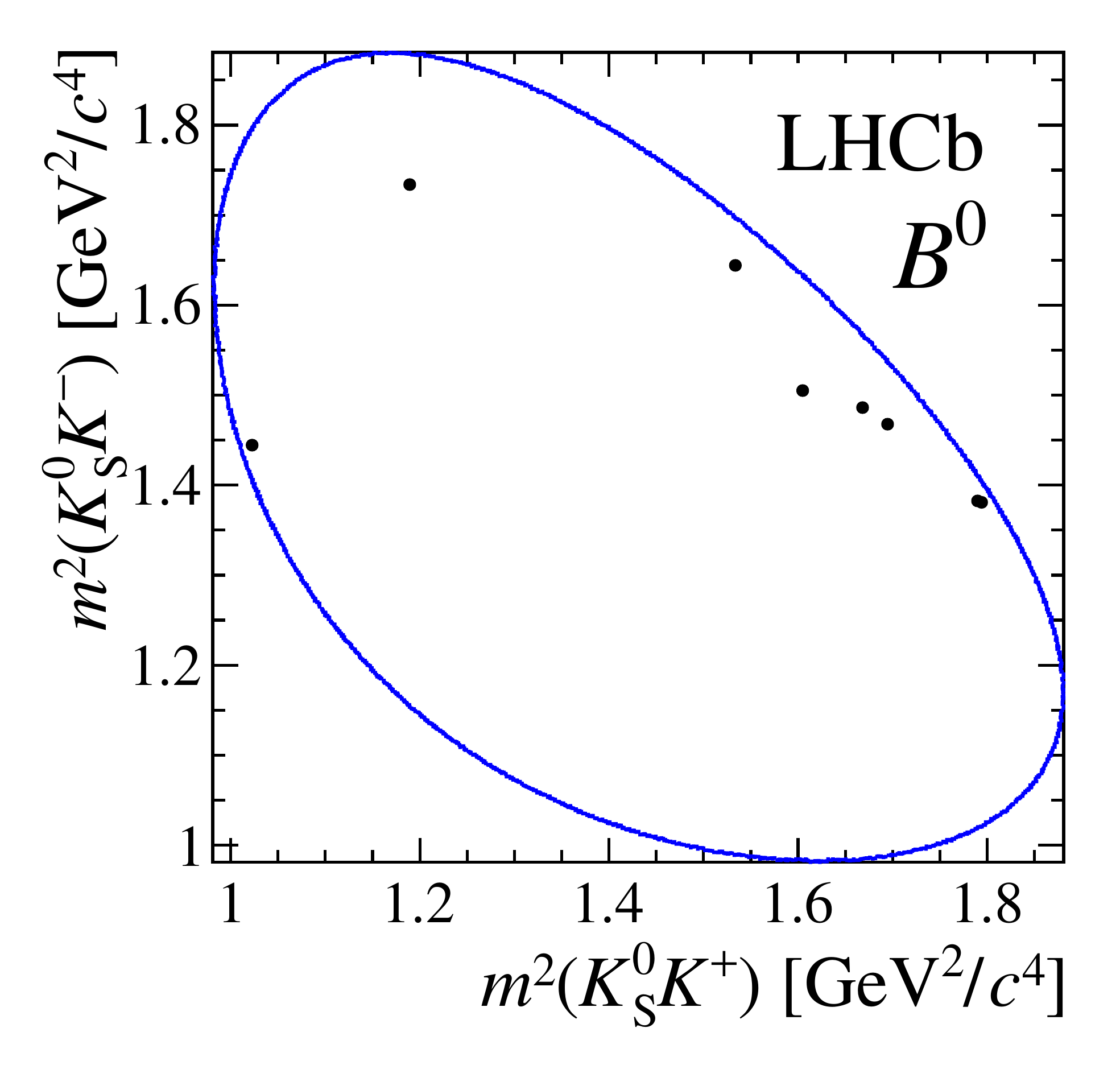}
\caption{Dalitz plots of candidates in the signal region for \DtoKsKK decays from (left) $\Bzb\to D\Kstarzb$ and (right) $\Bz\to D\Kstarz$ decays. The solid blue line indicates the kinematic boundary.}
\label{fig:dalitzkk}
\end{figure}

The Dalitz plots for \BtoDKst candidates restricted to the signal region for the two \DtoKshh final states are shown in Figs.~\ref{fig:dalitzpipi} and~\ref{fig:dalitzkk}. Separate plots are shown for \Bz and \Bzb decays.

\section{\boldmath Event selection and yield determination for \texorpdfstring{\BztoDstmu}{B0 -> D* mu nu} decays}

\label{sec:dmu}
A sample of \BztoDstmu, $\Dstarm \to \Dzb\pim$, $\Dzb\to\Kshh$ decays is used to determine the quantities $F_i$, defined in Eq.~(\ref{eq:fi}), as the expected fractions of \Dz decays falling into Dalitz plot bin $i$, taking into account the efficiency profile of the signal decay.
The semileptonic decay of the $B$ meson and the strong-interaction decay of the \Dstarpm meson allow the flavour of the \Dz meson to be determined from the charge of the muon and \Dstarpm daughter pion. This particular decay chain, involving a flavour-tagged \Dz decay, is chosen due to its high yield, low background level, and low mistag probability. The selection requirements are chosen to minimise changes to the efficiency profile with respect to that associated with the \BtoDKst channel and are the same as those listed in Ref.~\cite{LHCb-PAPER-2014-041}, with two exceptions. First, only events which pass the hardware trigger that selects muons with a transverse momentum $\pt>1.48\gevc$ are used. Those where the hardware trigger only satisfies the criterion of a high transverse energy deposit in the hadronic calorimeter are not considered. 
Second, the multivariate algorithm in the software trigger designed to select secondary vertices that are consistent with the decay of a \bquark hadron is identical to the one used for \BtoDKst candidates; an algorithm that also required the presence of a muon track was previously used.
The changes remove approximately $20\%$ of the sample used in Ref.~\cite{LHCb-PAPER-2014-041}; however, in simulated data they improve the agreement in the variation of the efficiency over the Dalitz plot between the \BtoDKst and \BztoDstmu decays.

\begin{figure}[tb]
\centering
\includegraphics[width=0.49\textwidth]{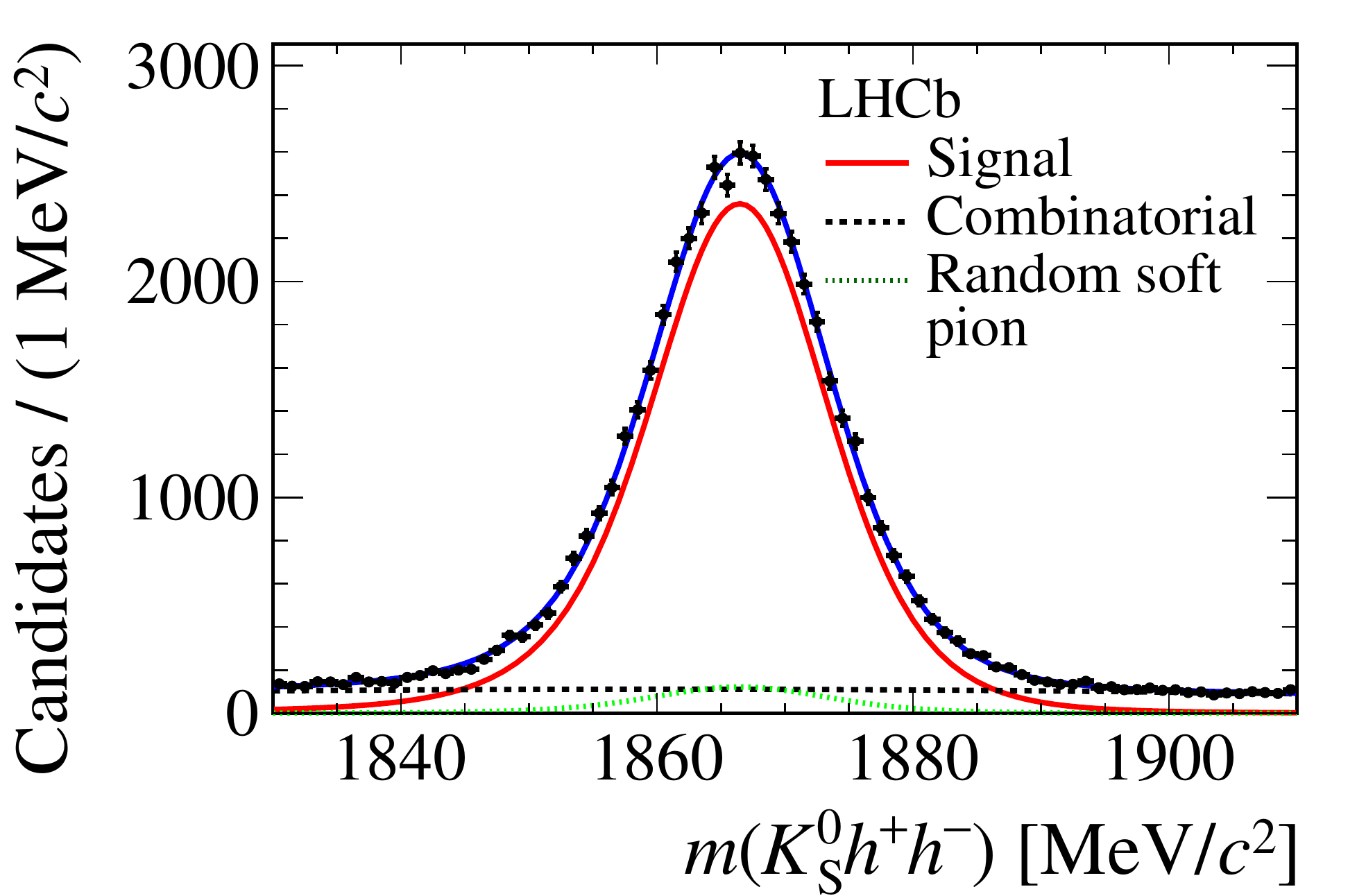}
\includegraphics[width=0.49\textwidth]{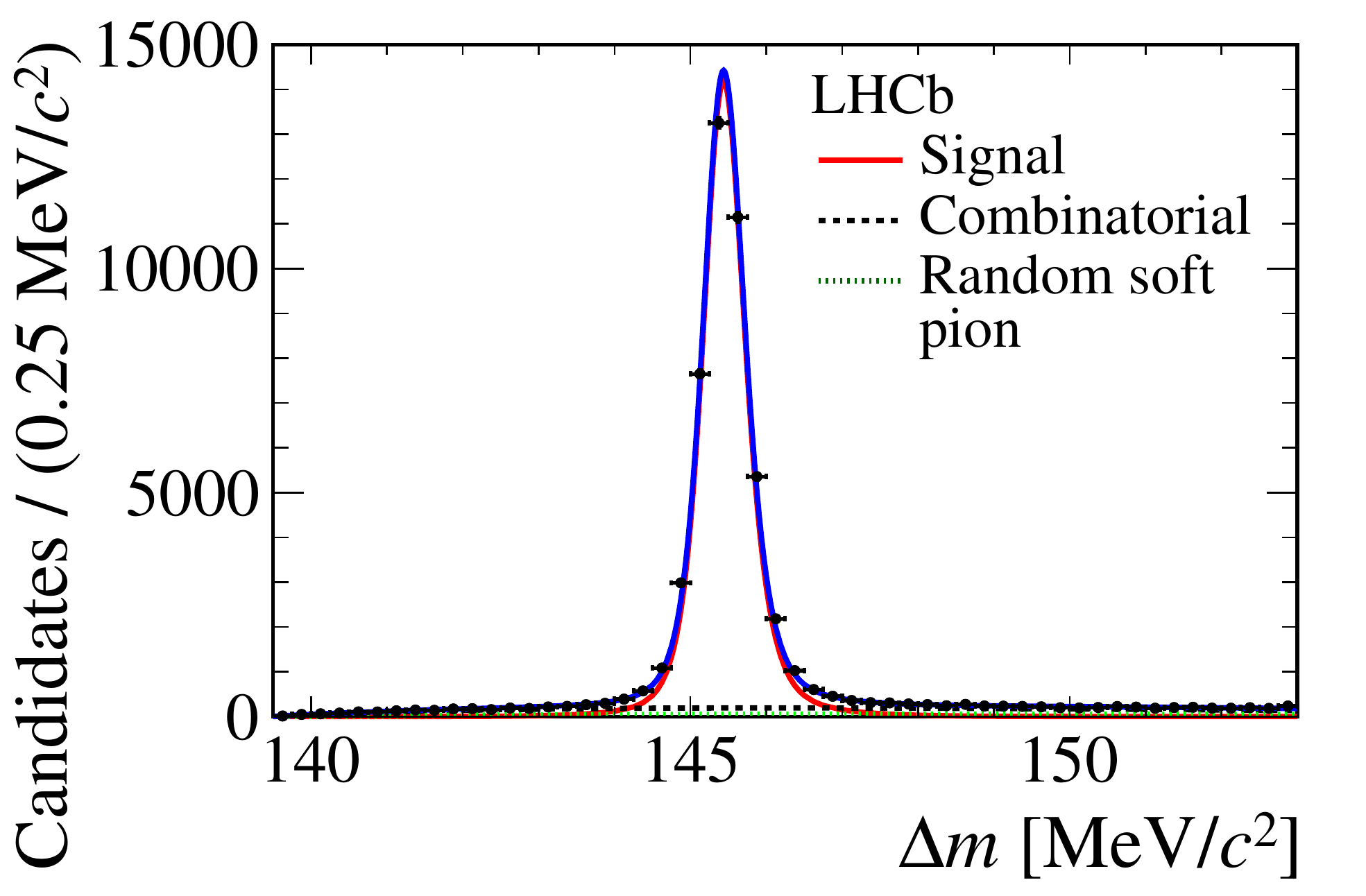}
\caption{Result of the simultaneous fit to \BztoDstmu, $\Dstarm\to\Dzb(\to\KsPiPi)\pim$ decays with downstream \KS candidates, in 2012 data. A two-dimensional fit is performed in (left) $m(\Kshh)$ and (right) $\Delta m$. The (blue) total fit PDF and the signal and background components are superimposed. }
\label{fig:slfit2012pipiDD}
\end{figure}

The \Dz invariant mass, $m(\Kshh)$, and the invariant mass difference $\Delta m \equiv m(\Kshh\pipm)-m(\Kshh)$ are fitted simultaneously to determine the signal and background yields. No significant correlation between these two variables is observed within the ranges chosen for the fit. 
This two-dimensional parametrisation allows the yield of selected candidates to be measured in three categories: true $\Dstarpm$ candidates (signal), candidates containing a true \Dz but a random soft pion (RSP) and candidates formed from random track combinations that fall within the fit range (combinatorial background). An example fit projection is shown in Fig.~\ref{fig:slfit2012pipiDD}. The result of the two-dimensional extended unbinned maximum likelihood fit is superimposed. The fit is performed simultaneously for the two \Dz final states and the two \KS categories, with some parameters allowed to vary between categories. Candidates selected from data recorded at $\sqs=7$ and $8\tev$ are fitted separately, due to their slightly different Dalitz plot efficiency profiles. The fit range is $1830 < m(\Kshh) < 1910\mevcc$ and $139.5 < \Delta m < 153.0\mevcc$. The PDFs used to model the various components in the fit are unchanged from those used in Ref.~\cite{LHCb-PAPER-2014-041}, where further details can be found.

A total signal yield of approximately 90\,000 (12\,000) \DtoKsPiPi (\DtoKsKK) candidates is obtained. The sample is three orders of magnitude larger than the \BtoDKst yield. The signal mass range is defined as 1840--$1890\mevcc$ (1850--$1880\mevcc$) in $m(\KsPiPi)$ ($m(\KsKK)$) and 143.9--$146.9\mevcc$ in $\Delta m$. Within this range the background contamination is 3--6\% depending on the category.

\begin{figure}[tb]
\centering
\includegraphics[width=0.48\textwidth]{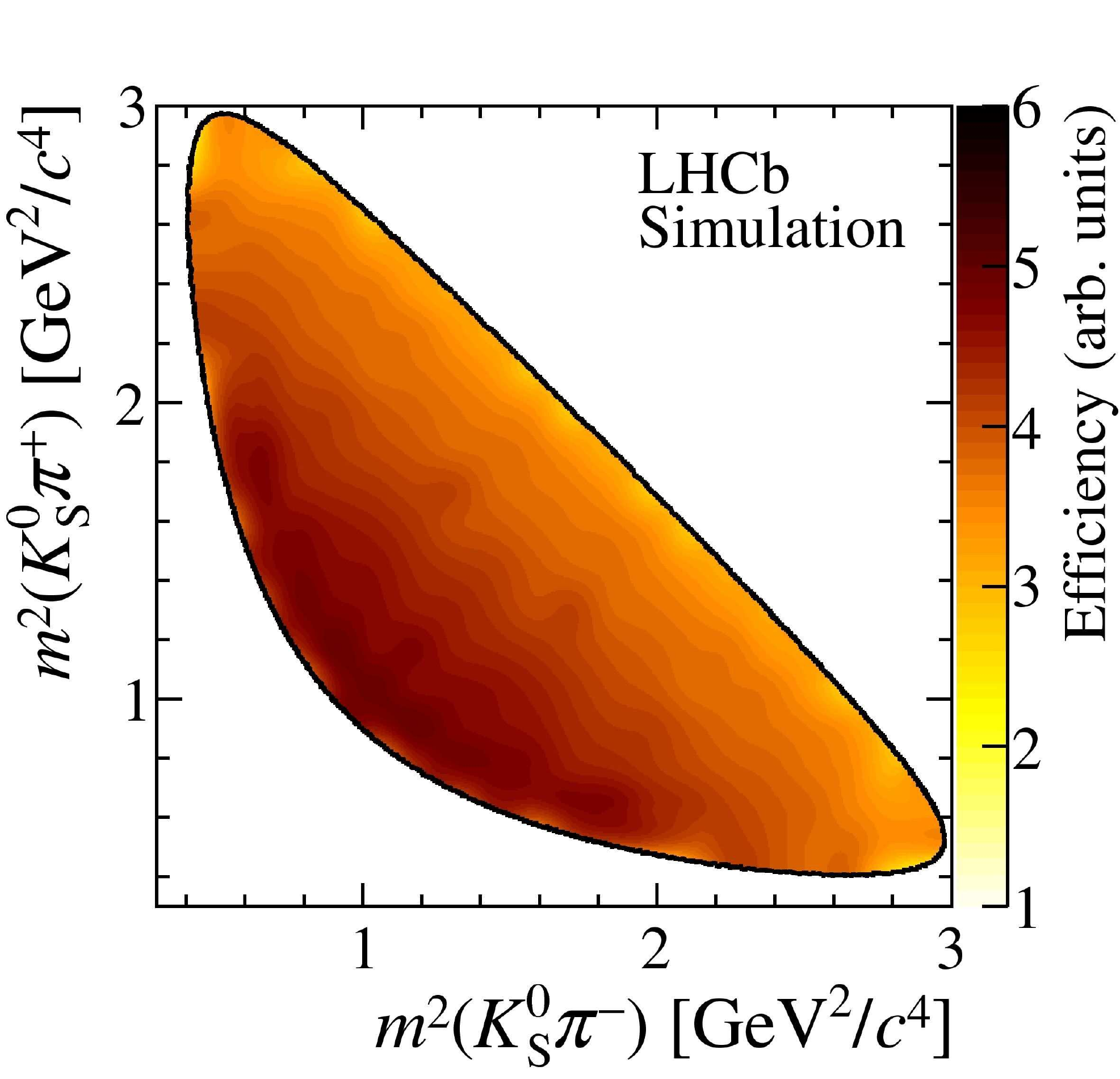}
\includegraphics[width=0.48\textwidth]{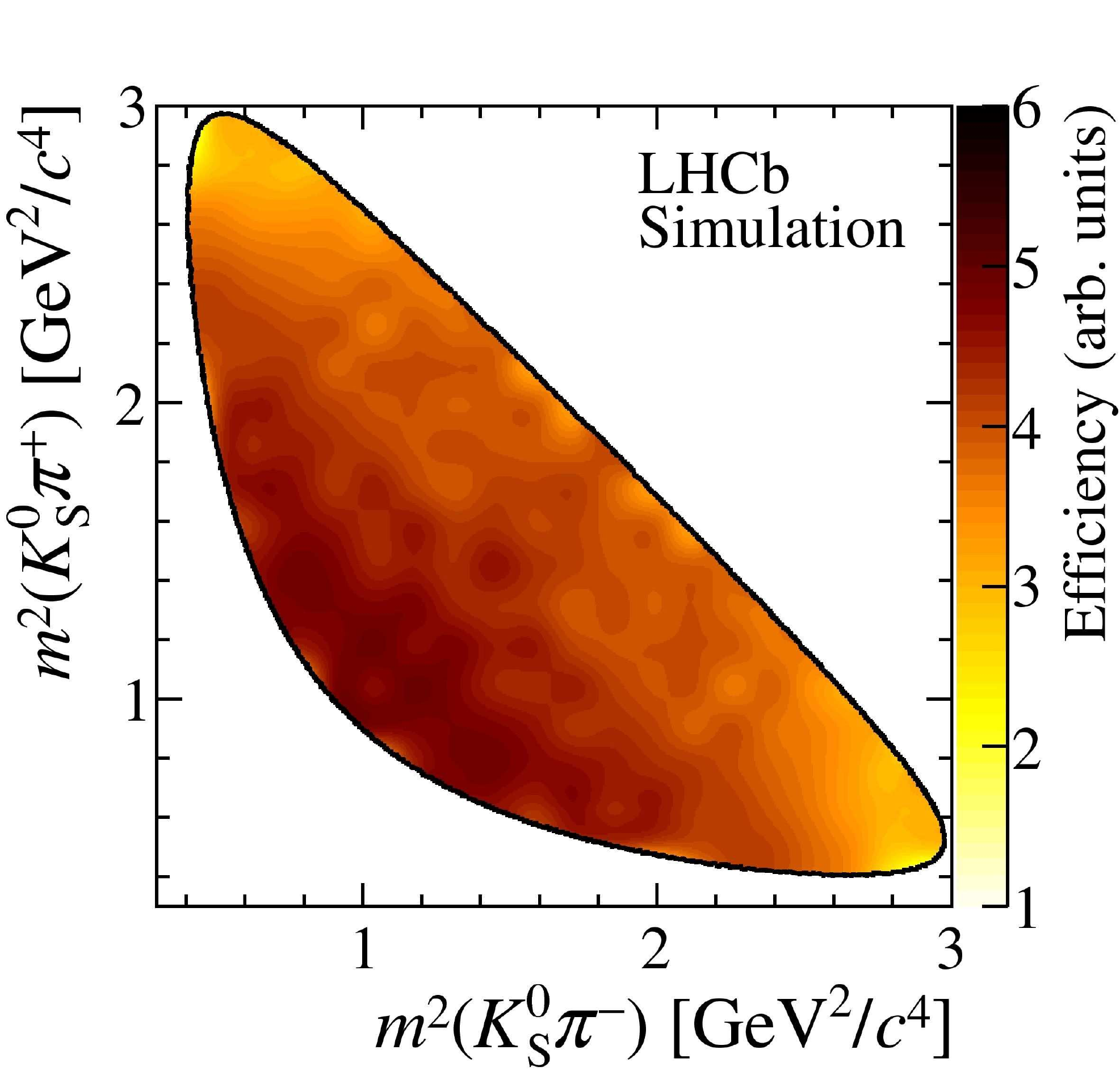} \\
\includegraphics[width=0.48\textwidth]{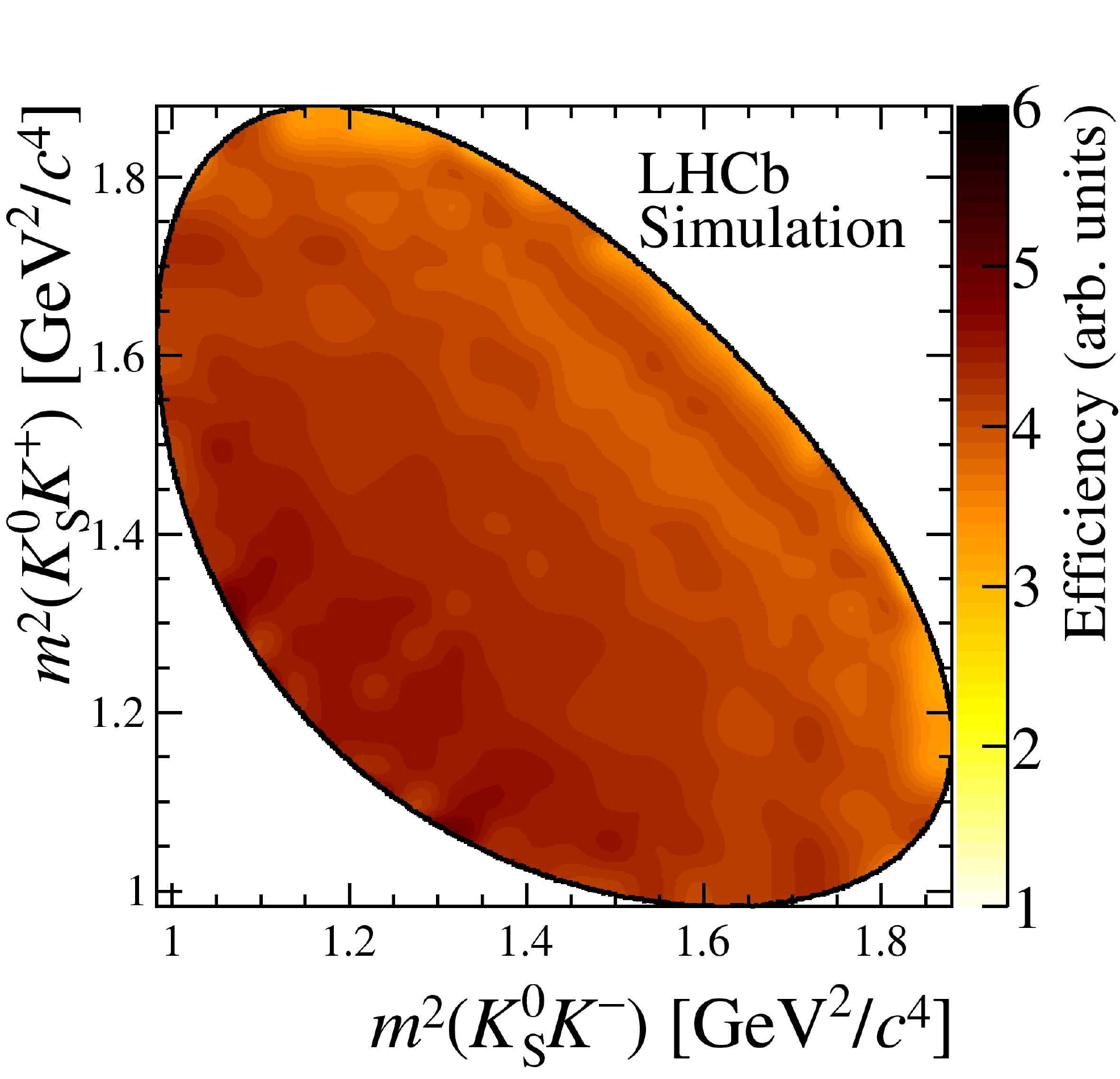}
\includegraphics[width=0.48\textwidth]{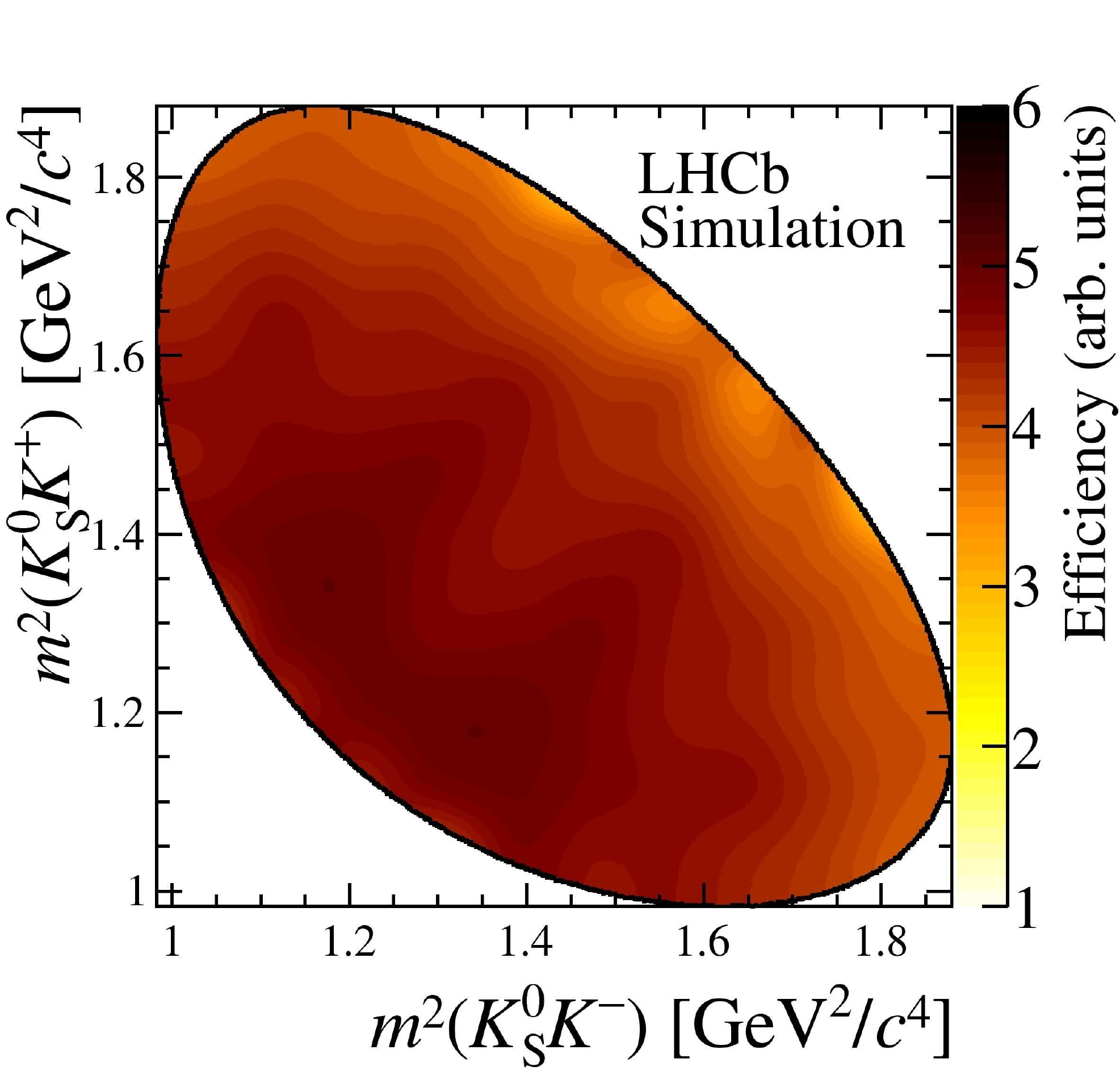}
\caption{Example efficiency profiles of (left) \BtoDKst and (right) \BztoDstmu decays in the simulation. The top (bottom) plots are for \DtoKsPiPi (\DtoKsKK) decays.}
\label{fig:mceff}
\end{figure}

The two-dimensional fit in $m(\Kshh)$ and $\Delta m$ of the \BztoDstmu decay is repeated in each Dalitz plot bin with all of the PDF parameters fixed, resulting in a raw control mode yield, $R_i$, for each bin $i$. The measured $R_i$ are not equivalent to the $F_i$ fractions required to determine the \CP parameters due to unavoidable differences from selection criteria in the efficiency profiles of the signal and control modes. Hence, a set of correction factors is determined from simulation. The efficiency profiles from simulation of \DtoKsPiPi decays are shown in Fig.~\ref{fig:mceff}. They show a variation of 50\% between the highest and lowest efficiency regions, although the efficiency changes within a bin are not as large. The variation over the \DtoKsKK Dalitz plot is smaller, at approximately 35\%.

The raw yields of the control decay must be corrected to take into account the differences in efficiency profiles. For each Dalitz plot bin a correction factor is determined,
\begin{equation}
\xi_{i} \equiv \frac{ \int_{i} d\msqmin \, d\msqplus \, |A_{D}(\msqmin,\msqplus)|^2\,\etaDKst(\msqmin,\msqplus)}{ \int_{i} d\msqmin \, d\msqplus \, |A_{D}(\msqmin,\msqplus)|^2\, \etaDst(\msqmin,\msqplus)},
\end{equation}
where \etaDKst and \etaDst are the efficiency profiles of the \BtoDKst and \BztoDstmu decays, respectively, and are determined with simulation. 
The amplitude models used to determine the Dalitz plot intensity for the correction factor are those from Ref.~\cite{BABAR2008} and Ref.~\cite{BABAR2010} for the \KsPiPi and \KsKK decays, respectively. The amplitude models used here only provide a description of the intensity distribution over the Dalitz plot and introduce no significant model dependence into the analysis. 
The correction factors are determined separately for data reconstructed with each \KS type, as the efficiency profile is different between the two \KS categories. This method of determining the $F_i$ parameters is preferable to using solely the amplitude models and \BtoDKst simulated events, since the method is data-driven and the efficiency correction causes deficiencies in the simulation and the model to cancel at first order.
The correction factors are within 10$\%$ of unity. The $F_i$ values can be determined via the relation $F_i = h'\xi_iR_i$, where $h'$ is a normalisation factor such that the sum of all $F_i$ is unity. The $F_i$ parameters are determined for each year of data taking and \KS category separately and are then combined in the fraction observed in the \BtoDKst signal region in data. 
The total uncertainty on $F_i$ is 5\% or less in all of the bins, and is a combination of the uncertainty on $R_i$ due to the size of the control channel, and the uncertainty on $\xi_i$ due to the limited size of the simulated samples. The two contributions are similar in size.

\section{\boldmath Dalitz plot fit to determine the \texorpdfstring{\CP}{CP}-violating parameters \texorpdfstring{$x_{\pm}$}{x} and \texorpdfstring{$y_{\pm}$}{y}}
\label{sec:dpfit}

The Dalitz plot fit is used to measure the \CP-violating parameters $x_{\pm}$ and $y_{\pm}$, as introduced in Sect.~\ref{sec:principles}.  Following Eqs.~(\ref{eq:populations}) and (\ref{eq:populations2}), these parameters can be determined from the populations of the \Bz and \Bzb Dalitz plot bins, given the external information of the $c_i$ and $s_i$ parameters from CLEO-c data, the values of $F_i$ from the semileptonic control decay modes and the measured value of $\kappa$. 
 
Although the absolute numbers of \Bz and \Bzb decays integrated over the $D$ Dalitz plot have some dependence on $x_{\pm}$ and $y_{\pm}$, the sensitivity gained compared to using just the relations in Eqs.~(\ref{eq:populations}) and (\ref{eq:populations2}) is negligible~\cite{Gershon}. Consequently, as stated previously, the integrated yields are not used and the analysis is insensitive to $B$ meson production and detection asymmetries. 

The \BtoDKst data are split into four categories, one for each $D$ decay and then by the charge of the \Kstarz daughter kaon. As in the case of the fit to the invariant mass, data from the two \KS categories are merged.
Each category is then divided into the Dalitz plot bins shown in Fig.~\ref{fig:bins}, where there are 16 bins for \DtoKsPiPi and 4 bins for \DtoKsKK. Since the Dalitz plots for \Bz and \Bzb data are analysed separately, this gives a total of 40 bins. The PDF parameters for the signal and background invariant mass distributions are fixed to the values determined in the invariant mass fit described in Sect.~\ref{sec:massfit}. 

The yield of the combinatorial background in each bin is a free parameter, apart from the yields in bins in which an auxiliary fit determines it to be negligible. It is necessary to set these to zero to facilitate the calculation of the covariance matrix. The total yield of \BstoDKst decays integrated over the Dalitz plot for each category is a free parameter.
The value of $r_B(\BstoDKst)$ is expected to be an order of magnitude smaller than $r_\Bz$ due to suppression from CKM factors. Hence, the fractions in each Dalitz plot bin are assigned assuming that \CP violation in these decays are negligible, which is also consistent with observations in Ref.~\cite{LHCb-PAPER-2015-059}. Therefore, the decay of the \Bs (\Bsb) meson contains a \Dzb (\Dz) meson. 
It is verified in simulation that the reconstruction efficiency over the $D$ Dalitz plot does not depend on the parent $B$ decay and hence the yield of \BstoDKst decays in bin $i$ is given by the relevant total yield multiplied by $F_{-i}$. 

The total yields of the \BstoDstKst, \BtoDrho and \BtoDK backgrounds in each category are determined by multiplying the total yield of \BstoDKst in that category by the values of $R_s$, $R_{\rho}$ and $R_{DK}$, respectively, that are listed in Table~\ref{tab:massfit:results}. The following assumptions are made about the Dalitz plot distributions of these backgrounds.
The \CP violation in \BstoDstKst decays is expected to be negligible as the underlying CKM factors are the same as that for \BstoDKst decays. Hence, the \BstoDstKst decays are distributed over the \DtoKshh Dalitz plot in the same way as \BstoDKst decays. The $D$ meson from \BtoDrho decays is assumed to be an equal admixture of \Dz and \Dzb and hence the yield is distributed according to ($F_{+i} + F_{-i}$), because the pion misidentified as a kaon is equally likely to be of either charge. In the case of the \BtoDK decay, \CP violation is expected and the yield is distributed according to Eqs.~(\ref{eq:populations}) and (\ref{eq:populations2}), where the values of the \CP violating parameters are those determined in Ref.~\cite{LHCb-PAPER-2014-041}.

The \BtoDKst yield in each bin is determined using the total yield of \BtoDKst in each category, which is a free parameter, and Eqs.~(\ref{eq:populations}) and (\ref{eq:populations2}). The parameters of interest, $x_{\pm}$ and $y_{\pm}$, are allowed to vary. The values of $c_i$ and $s_i$ are constrained to their measured values from CLEO~\cite{CLEOCISI}, assuming Gaussian errors and taking into account statistical and systematic correlations. The values of $F_i$ are fixed. The value of $\kappa$ is also fixed in the fit to the central value measured in Ref.~\cite{LHCb-PAPER-2015-059}.

An ensemble of 10\,000 pseudoexperiments is generated to validate the fit procedure.  In each pseudoexperiment the numbers and distributions of signal and background candidates are generated according to the expected distribution in data, taking care to smear the input values of $c_i$ and $s_i$. The full fit procedure is then performed. A variety of $x_{\pm}$ and $y_{\pm}$ values consistent with previous measurements is used~\cite{HFAG}. Small biases in the central values, with magnitudes around $10\%$ of the statistical uncertainty, are observed in the pseudoexperiments. These biases are due to the low event yields in some of the bins and they reduce in simulated experiments with higher yields. The central values are corrected for the biases.

\begin{figure}[tb]
\centering
\includegraphics[width=0.6\textwidth]{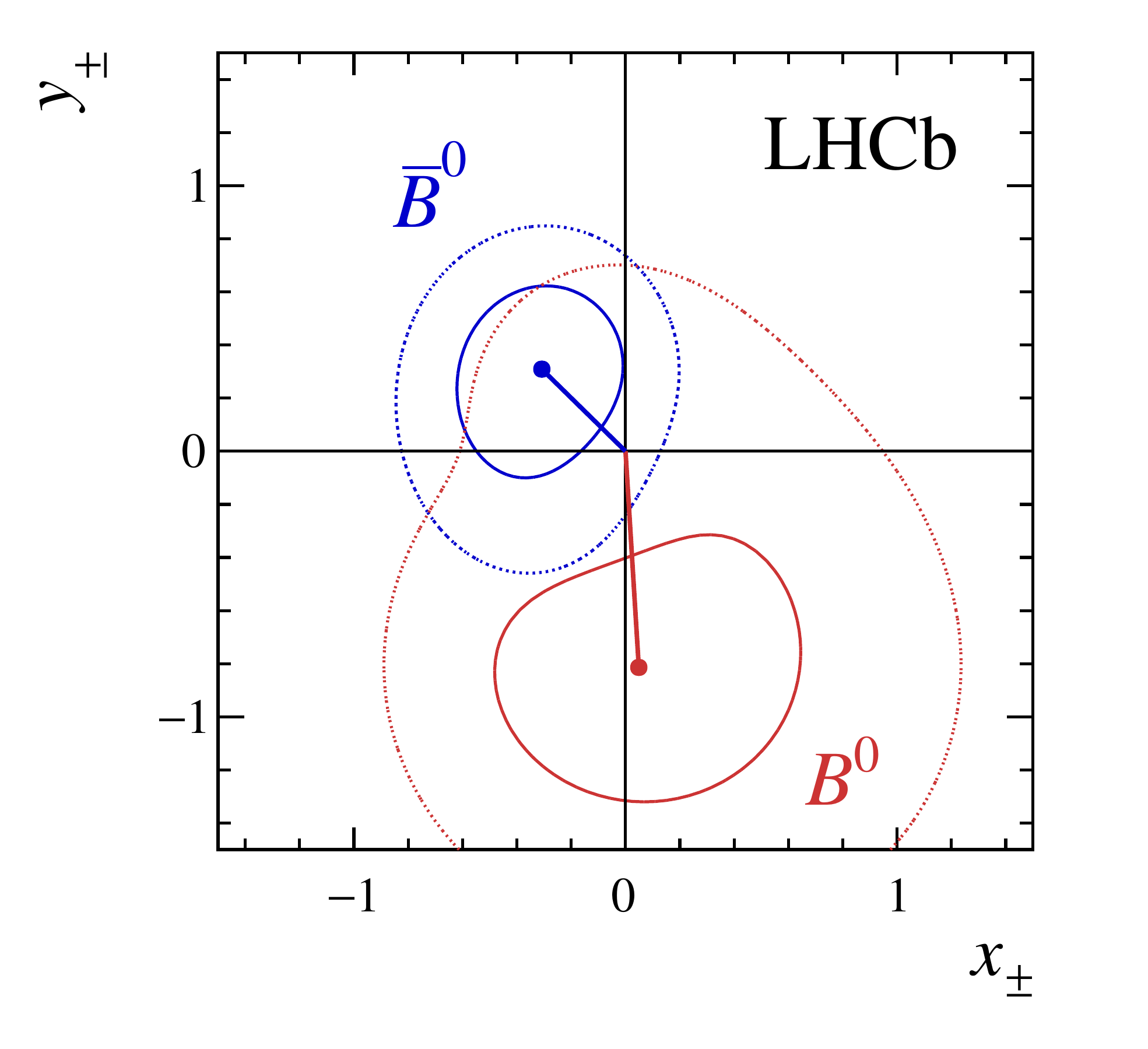}
\caption{Confidence levels at (solid) 68.3\% and (dotted) 95.5\% for (red, light)
$(x_+,y_+)$ and (blue, dark) $(x_{-}, y_{-})$ as measured in \BtoDKst decays (statistical uncertainties only). The parameters $(x_+,y_+)$ relate to \Bz decays and $(x_{-}, y_{-})$ refer to \Bzb decays. The points represent the best fit values.}
\label{fig:sunnysideup}
\end{figure}

The results of the fit are $\xp = 0.05 \pm 0.35$, $\xm = -0.31 \pm 0.20$, $\yp = -0.81\pm 0.28$, and $\ym = 0.31 \pm 0.21$. The statistical uncertainties are compatible with those predicted by the pseudoexperiments. 
The measured values of $(x_{\pm}, y_{\pm})$ from the fit to data, with their likelihood contours, corresponding to statistical uncertainties only, are displayed in Fig.~\ref{fig:sunnysideup}. 
The expected signature for a sample that exhibits \CP violation is that the two vectors defined by the coordinates $(x_-,y_-)$ and $(x_+,y_+)$ should both be non-zero in magnitude and have a non-zero opening angle. This opening angle is equal to $2\gamma$.
No evidence for \CP violation is observed. 

To investigate whether the binned fit gives an adequate description of the distribution of events over the Dalitz plot, the signal yield in each bin is fitted directly as a cross-check. A comparison of these yields and those predicted by the fitted values of \xpm and \ypm shows good agreement.

\section{Systematic uncertainties}
\label{sec:syst}

Systematic uncertainties are evaluated on the measurements of the Cartesian parameters and are presented in Table~\ref{tab:syst_all}. The source of each systematic uncertainty is described in turn below. Unless otherwise described, the systematic uncertainties are determined from an ensemble of pseudoexperiments where the simulated data are generated in an alternative configuration, and fitted with the default method described in Sect.~\ref{sec:dpfit}. The mean shift in the fitted values of \xpm and \ypm in comparison to their input values is taken as the systematic uncertainty. Uncertainties arising from the CLEO measurements are included within the statistical uncertainties since the values of $c_i$ and $s_i$ are constrained in the Dalitz plot fit. Their contribution to the statistical uncertainty is approximately 0.02 for \xpm and 0.05 for \ypm. 

\begin{table}[tbp]
\begin{center}
\caption{Summary of the systematic uncertainties for the parameters \xy. The various sources of systematic uncertainties are described in the main text.
}
\label{tab:syst_all}
\vspace{2mm}
\begin{tabular}{l c c c c}
Source & $\sigma(\xp)$ & $\sigma(\xm)$ & $\sigma(\yp)$ & $\sigma(\ym)$ \\
\hline
Efficiency corrections 			        & 0.019 & 0.034 & 0.021 & 0.005 \\
Efficiency combination 			        & 0.007 & 0.001 & 0.007 & 0.008 \\
Mass fit: $\alpha$ 		              & 0.002 & 0.005 & 0.021 & 0.020 \\
\phantom{Mass fit:} \BstoDstKst   	& 0.002 & 0.002 & 0.010 & 0.005 \\
\phantom{Mass fit:} \BtoDrho        & 0.002 & 0.003 & 0.004 & 0.001 \\
\phantom{Mass fit:} \BtoDK 						        & 0.000 & 0.000 & 0.000 & 0.000 \\
\phantom{Mass fit:} Signal shape			        & 0.005 & 0.003 & 0.003 & 0.002 \\
\phantom{Mass fit:} $\Bz \to \Dstarz\Kstarz$ 	& 0.006 & 0.007 & 0.008 & 0.004 \\
\phantom{Mass fit:} $B \to \Dstarz h$         & 0.001 & 0.001 & 0.007 & 0.005 \\
\phantom{Mass fit:} $B \to D\pi\pi\pi$        & 0.001 & 0.002 & 0.001 & 0.003 \\
Dalitz plot migration				        & 0.003 & 0.004 & 0.007 & 0.003 \\
Value of $\kappa$                   & 0.001 & 0.011 & 0.008 & 0.002 \\
Fitter bias                         & 0.004 & 0.014 & 0.042 & 0.042 \\ 
\hline
Total systematic 				            & 0.022 & 0.040 & 0.056 & 0.048 \\
\end{tabular}
\end{center}
\end{table}

A systematic uncertainty arises from imperfect modelling in the simulation used to derive the efficiency correction in the determination of the $F_i$ parameters. To determine this systematic uncertainty, a conservative approach is used, where an alternative set of $F_i$ values is determined using only the amplitude models and simulated \BtoDKst decays. These alternative $F_i$ are used in the generation of pseudoexperiments to determine the systematic uncertainty. A further uncertainty on the $F_i$ parameters arises from the fractions in which the individual $F_i$ parameters from the differing categories (year of data taking and \KS type) are combined. A second alternate set of $F_i$ are obtained by combining the values of $F_i$ for each category using the fractions of data observed in the \Bs mass window. The fractions in the \Bz window are statistically consistent with those observed in the \Bs mass window. The associated uncertainty is determined through the use of pseudoexperiments which are generated with the alternate set of $F_i$ values. 

Several systematic uncertainties are associated with the parametrisation of the invariant mass distribution. These arise from uncertainties in the shape of the \BstoDstKst background, the size of the \BtoDrho background, \CP violation in the \BtoDK background, the PDF shape used to describe the signal peak and the inclusion of backgrounds that are neglected in the nominal fit, because of their small yield. 

The uncertainty in the shape of the \BstoDstKst background arises from the relative contribution of the different \Dstarz decay and helicity state components, each of which have a different $DK\pi$ invariant mass distribution. 
A different parametrisation of the data with the lower mass limit extending down to 4900\mevcc results in a measurement $\alpha = 0.9 \pm 0.1$, in comparison to the value of $0.74\pm0.13$ obtained in the fit described in Sect.~\ref{sec:massfit}. 
Accounting for the difference in mass range, the uncertainty is estimated by generating pseudoexperiments with $\alpha=0.91$, and is found to be 2\e{-3} or less in each of the \CP parameters.

A separate systematic uncertainty is evaluated for the relative fraction of $\Dstarz \to \Dz\piz$ and $\Dstarz\to\Dz\gamma$ decays in the \BstoDstKst contribution. The uncertainties in the relative fractions are due to uncertainties in the branching fractions of the \Dstarz decays and in the selection efficiencies determined in simulation. In this case the systematic uncertainty is small and is determined by fitting the data repeatedly with the fractions smeared around the central values. 

The estimation of the \BtoDrho yield ignores the $\Bz \to D \pip \pim$ S-wave contributions, which will contribute if the misidentified $\pip\pim$ invariant mass falls within the \Kstarz mass window. The amplitude analysis of $\Bz\to D \pip\pim$ decays in Ref.~\cite{LHCb-PAPER-2014-070} is used to determine that the potential size of the S-wave contribution could increase the apparent \BtoDrho yield by approximately 50\%. Assuming that the additional S-wave contribution will have the same $DK\pi$ invariant mass distribution, the systematic uncertainty on the \CP parameters is estimated by generating pseudoexperiments with the \BtoDrho contribution increased by 50\%. The resulting uncertainties on \xy are lower than 4\e{-3}.

In the default fit the \CP parameters of the \BtoDK background are fixed to the central values measured in Ref.~\cite{LHCb-PAPER-2014-041}. The fits to the data are repeated with multiple values of the \CP parameters of the \BtoDK decay, smeared according to the measured uncertainties and correlations, and the shifts in \xy are found to be less than 0.001.

An alternative PDF to describe the \Bz and \Bs signals is considered by taking the sum of three Gaussian functions. The mean and width of the primary Gaussian is determined by performing a mass fit to data with the relative means and widths of the two secondary Gaussians taken from simulation. The systematic uncertainty is small and is estimated by generating pseudoexperiments with this alternative PDF.

In the default mass fit the contributions of \BtoDstKst, $B^\pm\to\Dstarz\pi^\pm$, $B^\pm\to\Dstarz K^\pm$ and $B^\pm \to D\pipm\pip\pim$ decays are ignored as they are estimated to contribute approximately $2-3$ events each. A systematic uncertainty from neglecting each of these decays is evaluated. 
The \BtoDstKst decays can be described with the same PDFs as the \BstoDstKst decays but shifted by the $\Bs - \Bz$ mass difference.
The \B mass fit described in Sect.~\ref{sec:massfit} is performed with this background included, where the yield of \BtoDstKst decays is constrained relative to that of the \BstoDKst in a similar manner to the \BstoDstKst decays. Although the addition of this background only has a small impact on the mass fit parameters, its \CP parameters are unknown. Hence, pseudoexperiments are generated with the \BtoDstKst background in three different \CP violating hypotheses and are fitted with the default configuration. The uncertainty is found to be less than 0.01 for all choices of the \CP parameters.
Further pseudoexperiments are generated with $B^\pm\to\Dstarz h^\pm$ and $B^\pm \to D\pipm\pip\pim$ decays, where their PDF shapes and yields are determined from simulation. Fitting the pseudoexperiments with the nominal fit demonstrates that the uncertainty due to ignoring these decays is 7\e{-3} or less for all \CP parameters.

The systematic uncertainty from the effect of candidates being assigned the wrong Dalitz plot bin number is considered. This can occur if reconstruction effects cause shifts in the measured values of $m^2_+$ and $m^2_-$ away from their true values. 
For both \BtoDKst and \BztoDstmu decays the resolution in $m^2_+$ and $m^2_-$ is approximately $0.005\gevgevcccc$ ($0.006\gevgevcccc$) for candidates with long (downstream) \KS decays. This is small compared to the typical width of a bin, but net migration can occur if the candidate lies close to the edge of a Dalitz plot bin. To first order, this effect is accounted for by use of the control channel, but residual effects enter due to the non-zero value of $r_\Bz$ in the signal decay, causing a different distribution in the Dalitz plot. The uncertainty due to these residual effects is determined via pseudoexperiments, in which different input $F_i$ values are used to reflect the residual migration. The size of this possible bias is found to vary between 3\e{-3} and 7\e{-3}.

The value of $\kappa$ has an associated uncertainty, and so pseudoexperiments are generated assuming the value $\kappa=0.912$, which corresponds to the central value of $\kappa$ lowered by one standard deviation. The mean shifts in \xy are of order 0.01. 
As described in Sect.~\ref{sec:dpfit}, the central values of the fit parameters \xpm and \ypm are corrected by a fitter bias that is determined with pseudoexperiments. The systematic uncertainty is assigned using half the size of the correction. 

The total experimental systematic uncertainty is determined by adding all sources in quadrature and is 0.02 on $x_+$, 0.04 on $x_-$, 0.06 on $y_+$, and 0.05 on $y_-$. These uncertainties are dominated by the efficiency corrections in $F_i$ and the fitter bias. The systematic uncertainties are less than 20\% of the corresponding statistical uncertainties.

\section{Results and interpretation}
\label{sec:Results}

The results for \xpm and \ypm are
\begin{align*}
\xp &= \phantom{-}0.05 \pm 0.35 \pm 0.02, \\ 
\xm &=          - 0.31 \pm 0.20 \pm 0.04, \\ 
\yp &=          - 0.81 \pm 0.28 \pm 0.06, \\ 
\ym &= \phantom{-}0.31 \pm 0.21 \pm 0.05,  
\end{align*}
where the first uncertainties are statistical and the second are systematic.
After accounting for all sources of uncertainty, the correlation matrix between the measured \xpm, \ypm parameters for the full data set is obtained, and is given in Table~\ref{tab:corrmatall}. Correlations for the statistical uncertainties are determined by the fit. The systematic uncertainties are only weakly correlated and the correlations are ignored.

\begin{table}[b]
\centering
\caption{Total correlation matrix, including statistical and systematic uncertainties, between the \xpm, \ypm parameters used in the extraction of $\gamma$. \label{tab:corrmatall}}
\begin{tabular}{c | c c c c }
      & $x_+$ & $x_-$ & $y_+$ & $y_-$ \\
\hline
$x_+$ & $\phantom{-} 1.00$ & $\phantom{-} 0.00$ & $\phantom{-} 0.13$ & $-0.01$ \\
$x_-$ & & $\phantom{-} 1.00$ & $-0.01$ & $\phantom{-} 0.14$ \\
$y_+$ & & & $\phantom{-} 1.00$ & $\phantom{-}0.02$ \\
$y_-$ & & & & $\phantom{-} 1.00$ \\
\end{tabular}
\end{table}

The results for $x_\pm$ and $y_\pm$ can be interpreted in terms of the underlying physics parameters $\gamma$, $r_{\Bz}$ and $\delta_{\Bz}$.  This interpretation is performed using a Neyman construction with Feldman-Cousins ordering~\cite{FELDMANCOUSINS}, using the same procedure as described in Ref.~\cite{BELLEMODIND}, yielding confidence levels for the three physics parameters. 

\begin{figure}[tb]
\centering
\includegraphics[width=0.90\textwidth]{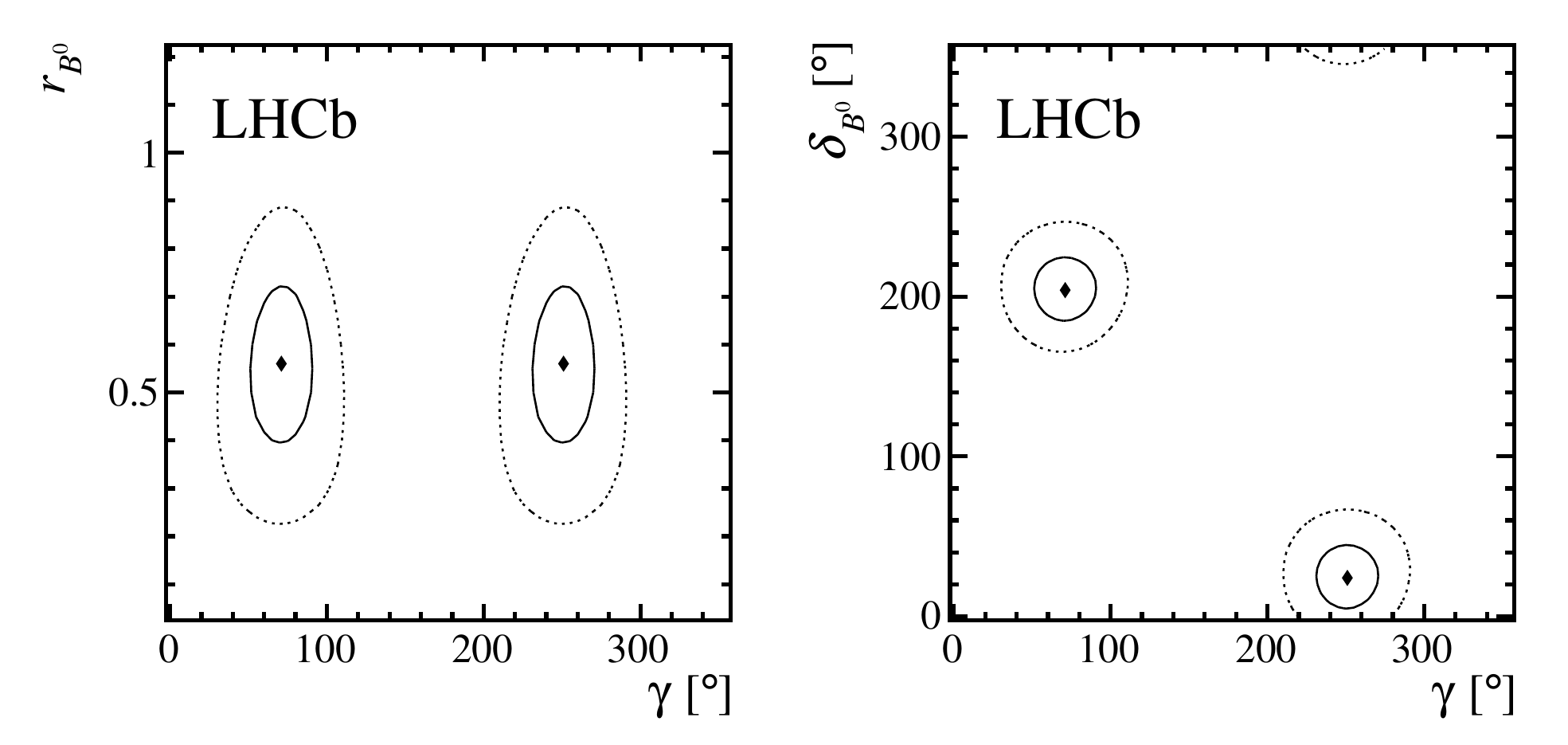}
\caption{The three-dimensional confidence volumes projected onto the $(\gamma, r_{\Bz})$ and $(\gamma, \delta_{\Bz})$ planes. The confidence levels correspond to 68.3\% and 95.5\% confidence levels when projected onto one dimension and are denoted by solid and dotted contours, respectively. The diamonds mark the central values.}
\label{fig:twodscans}
\end{figure}

In Fig.~\ref{fig:twodscans}, the projections of the three-dimensional surfaces containing the one and two standard deviation volumes (\ie, $\Delta\chisq=1$ and 4) onto the $(\gamma, r_{\Bz})$ and $(\gamma, \delta_{\Bz})$ planes are shown; the statistical and systematic uncertainties on $x_\pm$ and $y_\pm$ are combined in quadrature.
The solution for the physics parameters has a two-fold ambiguity, with a second solution corresponding to $(\gamma, \delta_{\Bz}) \rightarrow (\gamma + 180\degrees, \delta_{\Bz} + 180\degrees)$.  For the solution that satisfies $0 < \gamma < 180\degrees$, the following results are obtained: 
\begin{align*}
r_{\Bz} &= 0.56\pm 0.17,\\
\delta_{\Bz} &= (204\,^{+21}_{-20})\degrees, \\ 
\gamma  &= (71\pm 20)\degrees.
\end{align*}
The central value for $\gamma$ is consistent with the world average from previous measurements~\cite{CKMFitter,UTFit}. The value for $r_{\Bz}$, while consistent with current knowledge, has a central value that is larger than expected~\cite{bab2body,belle2body,BGGSZ,BABAR2008}.
The results are also consistent with, but cannot be combined with, the model-dependent analysis of the same dataset performed by LHCb~\cite{LHCb-PAPER-2016-007}.

A key advantage of having direct measurements of \xpm and \ypm is that there is only a two-fold ambiguity in the value of $\gamma$ from the trigonometric expressions. This means that when combined with the results of other \CP violation studies in \BtoDKst decays such as those in Ref.~\cite{LHCb-PAPER-2014-028}, these measurements will provide strong constraints on the hadronic parameters, and will provide improved sensitivity to $\gamma$ when combined with all other measurements.

\section*{Acknowledgements}

\noindent We express our gratitude to our colleagues in the CERN
accelerator departments for the excellent performance of the LHC. We
thank the technical and administrative staff at the LHCb
institutes. We acknowledge support from CERN and from the national
agencies: CAPES, CNPq, FAPERJ and FINEP (Brazil); NSFC (China);
CNRS/IN2P3 (France); BMBF, DFG and MPG (Germany); INFN (Italy); 
FOM and NWO (The Netherlands); MNiSW and NCN (Poland); MEN/IFA (Romania); 
MinES and FANO (Russia); MinECo (Spain); SNSF and SER (Switzerland); 
NASU (Ukraine); STFC (United Kingdom); NSF (USA).
We acknowledge the computing resources that are provided by CERN, IN2P3 (France), KIT and DESY (Germany), INFN (Italy), SURF (The Netherlands), PIC (Spain), GridPP (United Kingdom), RRCKI and Yandex LLC (Russia), CSCS (Switzerland), IFIN-HH (Romania), CBPF (Brazil), PL-GRID (Poland) and OSC (USA). We are indebted to the communities behind the multiple open 
source software packages on which we depend.
Individual groups or members have received support from AvH Foundation (Germany),
EPLANET, Marie Sk\l{}odowska-Curie Actions and ERC (European Union), 
Conseil G\'{e}n\'{e}ral de Haute-Savoie, Labex ENIGMASS and OCEVU, 
R\'{e}gion Auvergne (France), RFBR and Yandex LLC (Russia), GVA, XuntaGal and GENCAT (Spain), Herchel Smith Fund, The Royal Society, Royal Commission for the Exhibition of 1851 and the Leverhulme Trust (United Kingdom).

\clearpage
\newpage
\addcontentsline{toc}{section}{References}
\setboolean{inbibliography}{true}
\bibliographystyle{LHCb}
\bibliography{main,LHCb-PAPER,LHCb-CONF,LHCb-DP,LHCb-TDR}


\newpage
\centerline{\large\bf LHCb collaboration}
\begin{flushleft}
\small
R.~Aaij$^{39}$, 
C.~Abell\'{a}n~Beteta$^{41}$, 
B.~Adeva$^{38}$, 
M.~Adinolfi$^{47}$, 
Z.~Ajaltouni$^{5}$, 
S.~Akar$^{6}$, 
J.~Albrecht$^{10}$, 
F.~Alessio$^{39}$, 
M.~Alexander$^{52}$, 
S.~Ali$^{42}$
% ...

{\footnotesize \it
% ...
$ ^{5}$Clermont Universit\'{e}, Universit\'{e} Blaise Pascal, CNRS/IN2P3, LPC, Clermont-Ferrand, France\\
$ ^{6}$CPPM, Aix-Marseille Universit\'{e}, CNRS/IN2P3, Marseille, France\\
% ...
$ ^{10}$Fakult\"{a}t Physik, Technische Universit\"{a}t Dortmund, Dortmund, Germany\\
% ...
$ ^{38}$Universidad de Santiago de Compostela, Santiago de Compostela, Spain\\
$ ^{39}$European Organization for Nuclear Research (CERN), Geneva, Switzerland\\
% ...
$ ^{41}$Physik-Institut, Universit\"{a}t Z\"{u}rich, Z\"{u}rich, Switzerland\\
$ ^{42}$Nikhef National Institute for Subatomic Physics, Amsterdam, The Netherlands\\
% ...
$ ^{47}$H.H. Wills Physics Laboratory, University of Bristol, Bristol, United Kingdom\\
% ...
$ ^{52}$School of Physics and Astronomy, University of Glasgow, Glasgow, United Kingdom\\
% ...
}
\end{flushleft}

\end{document}